\documentclass[traditabstract,10pt]{aa}

\usepackage[T1]{fontenc}
\usepackage{amsmath, amssymb, amsfonts, latexsym}
\usepackage[breaklinks,colorlinks,citecolor=blue]{hyperref} 
\usepackage{ifthen}
\usepackage[usenames,dvipsnames,table]{xcolor}
\usepackage{placeins}
\usepackage{txfonts}
\let\vec\mathbf
\usepackage{xspace}
\usepackage{soul}

\usepackage{graphicx}
\graphicspath{{Figures/}}
\usepackage{float}
\usepackage{epstopdf}
\usepackage[skip=5pt,belowskip=-3.5pt]{caption}
 
\usepackage[nonamebreak]{natbib}
\bibpunct{(}{)}{;}{a}{}{,}
\providecommand{\sorthelp}[1]{}

\newcommand{\npipe}{\texttt{NPIPE}\xspace}
\newcommand{\hillipop}{\texttt{HiLLiPoP}\xspace}
\newcommand{\lollipop}{\texttt{LoLLiPoP}\xspace}
\newcommand{\hlp}{hlp}
\newcommand{\lowlT}{{lowT}}
\newcommand{\lowlE}{{lolE}}

\newcommand{\quickpol}{\texttt{QuickPol}\xspace}
\newcommand{\commander}{\texttt{Commander}\xspace}
\newcommand{\camspec}{\texttt{CamSpec}\xspace}
\newcommand{\xpol}{\texttt{Xpol}\xspace}
\newcommand{\plik}{\texttt{Plik}\xspace}
\newcommand{\planck}{\textit{Planck}\xspace}
\newcommand{\lcdm}{$\Lambda$CDM\xspace}
\newcommand{\myurl}[1]{\href{http://#1}{\tt #1}}
\def\Alens{\ifmmode{A_\mathrm{L}}\else{$A_\mathrm{L}$}\fi}
\def\Omegak{\ifmmode{\Omega_{K}}\else{$\Omega_{K}$}\fi}
\def\Neff{\ifmmode{N_\mathrm{eff}}\else{$N_\mathrm{eff}$}\fi}
\def\mnu{\ifmmode{\textstyle \sum m_\nu}\else{$\textstyle \sum m_\nu$}\fi}
\newcommand{\nside}{\ifmmode {N_{\rm side}} \else $N_{\rm side}$ \fi}

\newcommand{\W}[4]{{W^{\scriptscriptstyle #1,#2}_{\scriptscriptstyle #3#4}}}
\newcommand{\Cl}[2]{C_{#1}^{#2}}
\newcommand{\alm}{\ifmmode {\vec{a}_{\ell m}} \else $\vec{a}_{\ell m}$\fi}
\def\VEV#1{{ \left\langle #1 \right\rangle }}
\newcommand{\lm}{{\ell m}}
\newcommand{\lmprime}{{\ell'm'}}
\newcommand{\lmun}{{\ell_1m_1}}

\newcommand{\lowl}{{low-$\ell$}}
\newcommand{\highl}{{high-$\ell$}}

\def\setsymbol#1#2{\expandafter\def\csname #1\endcsname{#2}}
\def\getsymbol#1{\csname #1\endcsname}

\def\Planck{\textit{Planck}}

\newbox\tablebox    \newdimen\tablewidth
\def\leaderfil{\leaders\hbox to 5pt{\hss.\hss}\hfil}
\def\tablenote#1 #2\par{\begingroup \parindent=0.8em
    \abovedisplayshortskip=0pt\belowdisplayshortskip=0pt
    \noindent
    $$\hss\vbox{\hsize\tablewidth \hangindent=\parindent \hangafter=1 \noindent
    \hbox to \parindent{$^#1$\hss}\strut#2\strut\par}\hss$$
    \endgroup}

\def\L2{\ifmmode L_2\else $L_2$\fi}

\def\DeltaT{\ifmmode \Delta T\else $\Delta T$\fi}
\def\deltat{\ifmmode \Delta t\else $\Delta t$\fi}
\def\fknee{\ifmmode f_{\rm knee}\else $f_{\rm knee}$\fi}
\def\Fmax{\ifmmode F_{\rm max}\else $F_{\rm max}$\fi}
\def\solar{\ifmmode{\rm M}_{\mathord\odot}\else${\rm M}_{\mathord\odot}$\fi}
\def\Msolar{\ifmmode{\rm M}_{\mathord\odot}\else${\rm M}_{\mathord\odot}$\fi}
\def\Lsolar{\ifmmode{\rm L}_{\mathord\odot}\else${\rm L}_{\mathord\odot}$\fi}
\def\inv{\ifmmode^{-1}\else$^{-1}$\fi}
\def\mo{\ifmmode^{-1}\else$^{-1}$\fi}
\def\sup#1{\ifmmode ^{\rm #1}\else $^{\rm #1}$\fi}
\def\expo#1{\ifmmode \times 10^{#1}\else $\times 10^{#1}$\fi}
\def\,{\thinspace}
\def\lsim{\mathrel{\raise .4ex\hbox{\rlap{$<$}\lower 1.2ex\hbox{$\sim$}}}}
\def\gsim{\mathrel{\raise .4ex\hbox{\rlap{$>$}\lower 1.2ex\hbox{$\sim$}}}}

\def\simprop{\mathrel{\raise .4ex\hbox{\rlap{$\propto$}\lower 1.2ex\hbox{$\sim$}}}}
\def\deg{\ifmmode^\circ\else$^\circ$\fi}
\def\pdeg{\ifmmode $\setbox0=\hbox{$^{\circ}$}\rlap{\hskip.11\wd0 .}$^{\circ}
          \else \setbox0=\hbox{$^{\circ}$}\rlap{\hskip.11\wd0 .}$^{\circ}$\fi}
\def\arcs{\ifmmode {^{\scriptstyle\prime\prime}}
          \else $^{\scriptstyle\prime\prime}$\fi}
\def\arcm{\ifmmode {^{\scriptstyle\prime}}
          \else $^{\scriptstyle\prime}$\fi}
\newdimen\sa  \newdimen\sb
\def\parcs{\sa=.07em \sb=.03em
     \ifmmode \hbox{\rlap{.}}^{\scriptstyle\prime\kern -\sb\prime}\hbox{\kern -\sa}
     \else \rlap{.}$^{\scriptstyle\prime\kern -\sb\prime}$\kern -\sa\fi}
\def\parcm{\sa=.08em \sb=.03em
     \ifmmode \hbox{\rlap{.}\kern\sa}^{\scriptstyle\prime}\hbox{\kern-\sb}
     \else \rlap{.}\kern\sa$^{\scriptstyle\prime}$\kern-\sb\fi}
\def\ra[#1 #2 #3.#4]{#1\sup{h}#2\sup{m}#3\sup{s}\llap.#4}
\def\dec[#1 #2 #3.#4]{#1\deg#2\arcm#3\arcs\llap.#4}
\def\deco[#1 #2 #3]{#1\deg#2\arcm#3\arcs}
\def\rra[#1 #2]{#1\sup{h}#2\sup{m}}

\def\dots{\relax\ifmmode \ldots\else $\ldots$\fi}
\def\WHzsr{\ifmmode $W\,Hz\mo\,sr\mo$\else W\,Hz\mo\,sr\mo\fi}
\def\mHz{\ifmmode $\,mHz$\else \,mHz\fi}
\def\GHz{\ifmmode $\,GHz$\else \,GHz\fi}
\def\mKs{\ifmmode $\,mK\,s$^{1/2}\else \,mK\,s$^{1/2}$\fi}
\def\muKs{\ifmmode \,\mu$K\,s$^{1/2}\else \,$\mu$K\,s$^{1/2}$\fi}
\def\muKRJs{\ifmmode \,\mu$K$_{\rm RJ}$\,s$^{1/2}\else \,$\mu$K$_{\rm RJ}$\,s$^{1/2}$\fi}
\def\muKHz{\ifmmode \,\mu$K\,Hz$^{-1/2}\else \,$\mu$K\,Hz$^{-1/2}$\fi}
\def\MJysr{\ifmmode \,$MJy\,sr\mo$\else \,MJy\,sr\mo\fi}
\def\MJysrmK{\ifmmode \,$MJy\,sr\mo$\,mK$_{\rm CMB}\mo\else \,MJy\,sr\mo\,mK$_{\rm CMB}\mo$\fi}
\def\microns{\ifmmode \,\mu$m$\else \,$\mu$m\fi}

\def\muK{\ifmmode \,\mu$K$\else \,$\mu$\hbox{K}\fi}
\def\microK{\ifmmode \,\mu$K$\else \,$\mu$\hbox{K}\fi}
\def\muW{\ifmmode \,\mu$W$\else \,$\mu$\hbox{W}\fi}
\def\kms{\ifmmode $\,km\,s$^{-1}\else \,km\,s$^{-1}$\fi}
\def\kmsMpc{\ifmmode $\,\kms\,Mpc\mo$\else \,\kms\,Mpc\mo\fi}
\providecommand{\sorthelp}[1]{}

\begin{document}

\title{Cosmological parameters derived from the final (PR4) \textit{Planck} data release}

\author{
M.~Tristram\inst{\ref{1}}
\and
A.~J.~Banday\inst{\ref{2}}
\and
M.~Douspis\inst{\ref{3}}
\and
X.~Garrido\inst{\ref{1}}
\and
K.~M.~G\'{o}rski\inst{\ref{4}, \ref{5}}
\and
S.~Henrot-Versill\'{e}\inst{\ref{1}}
\and
L.~T.~Hergt\inst{\ref{6}}
\and
S.~Ili\'{c}\inst{\ref{1},\ref{7}}
\and
R.~Keskitalo\inst{\ref{8},\ref{9}}
\and
G.~Lagache\inst{\ref{10}}
\and
C.~R.~Lawrence\inst{\ref{4}}
\and
B.~Partridge\inst{\ref{11}}
\and
D.~Scott\inst{\ref{6}}
}

\institute{
Universit\'{e} Paris-Saclay, CNRS/IN2P3, IJCLab, 91405 Orsay, France
\label{1}
\and
IRAP, Universit\'{e} de Toulouse, CNRS, CNES, UPS, (Toulouse), France
\label{2}
\and
Universit\'{e} Paris-Saclay, CNRS, Institut d'Astrophysique Spatiale, 91405, Orsay, France
\label{3}
\and
Jet Propulsion Laboratory, California Institute of Technology, 4800 Oak Grove Drive, Pasadena, California, U.S.A.
\label{4}
\and
Warsaw University Observatory, Aleje Ujazdowskie 4, 00-478 Warszawa, Poland
\label{5}
\and
Department of Physics \& Astronomy, University of British Columbia, 6224 Agricultural Road, Vancouver, British Columbia, V6T 1Z1, Canada
\label{6}
\and
Centre National d'Etudes Spatiales -- Centre Spatial de Toulouse, 18 Avenue Edouard Belin, 31401 Toulouse Cedex 9, France
\label{7}
\and
Computational Cosmology Center, Lawrence Berkeley National Laboratory, Berkeley, California, 94720, U.S.A.
\label{8}
\and
Space Sciences Laboratory, University of California, Berkeley, California, 94720, U.S.A.
\label{9}
\and
Aix Marseille Universit\'{e}, CNRS, CNES, LAM, Marseille, France
\label{10}
\and
Department of Astronomy, Haverford College, Haverford, Pennsylvania, 19041, U.S.A.
\label{11}
}

\abstract{We present constraints on cosmological parameters using maps from the last \planck data release (PR4). In particular, we detail an upgraded version of the cosmic microwave background likelihood, \hillipop, based on angular power spectra and relying on a physical modelling of the foreground residuals in the spectral domain. This new version of the likelihood retains a larger sky fraction (up to 75\,\%) and uses an extended multipole range. Using this likelihood, along with \lowl\ measurements from \lollipop, we derive constraints on \lcdm parameters that are in good agreement with previous \planck 2018 results, but with 10\,\% to 20\,\% smaller uncertainties.
We demonstrate that the foregrounds can be accurately described in the spectral domain with only negligible impact on \lcdm parameters.
We also derive constraints on single-parameter extensions to \lcdm including \Alens, \Omegak, \Neff, and \mnu. 
Noteworthy results from this updated analysis include a lensing amplitude value of $\Alens=1.039\pm0.052$, which aligns more closely with theoretical expectations within the \lcdm framework.
Additionally, our curvature measurement, $\Omegak = -0.012 \pm 0.010$, now demonstrates complete consistency with a flat universe, and our measurement of $S_8$ is closer to the measurements derived from large-scale structure surveys (at the 1.5\,$\sigma$ level). We also add constraints from PR4 lensing, making the combination the most constraining data set that is currently available from \Planck. Additionally we explore adding baryon acoustic oscillation data, which tightens limits on some particular extensions to the standard cosmology.}

\keywords{Cosmic background radiation -- cosmological parameters -- cosmology: observations -- methods: data analysis}

\date{\today}

\maketitle

\section{Introduction}

Since the first results were released in 2013, the \planck satellite's measurements of the cosmic microwave background (CMB) anisotropies have provided highly precise constraints on cosmological models. These measurements have tested the cosmological-constant-dominated cold dark matter (\lcdm) model, given tight constraints on its parameters, and ruled out many  plausible extensions. As a consequence, the best-fitting 6-parameter \lcdm model is now frequently used as the standard reference to be compared to new observational results, and when combining with other data sets to provide further constraints.

Since the last Planck Collaboration cosmological analysis in 2018 \citep{planck2016-l06}, the very last version of the \planck data processing, called \npipe, was released as the \planck Public Release 4 (PR4) and extensively detailed in \citet{planck2020-LVII}. As well as including previously neglected data from the repointing periods, \npipe processed the entire set of \planck channels within the same framework, including the latest versions of corrections for systematics and data treatment. 

In this paper, our objective is to enhance the precision on cosmological parameters through the utilization of PR4 data. Indeed, we expect better sensitivity on almost all cosmological parameters owing to improved map sensitivity. Additionally, we look for better internal consistency for the lensing amplitude affecting the primordial CMB anisotropies.
We thus derive constraints on cosmology using both \lowl\ and \highl\ likelihoods based on \planck PR4. The only part still relying on PR3 (also known as \textit{Planck} 2018) is the \lowl\ temperature likelihood, \commander, as we do not anticipate significant improvements at large scales in temperature between PR3 and PR4.
On the other hand, our analysis includes the large scales in polarization from PR4 for which the \npipe processing provides a significant improvement compared to PR3.

Since the foregrounds dominate polarization at large scales, for the \lowl\ likelihood, \lollipop, we make use of component-separated CMB maps processed by \commander using the whole range of \planck polarized frequencies from 30 to 353\GHz. This has been extensively discussed in \citet{tristram:2021} and \citet{tristram:2022}, where it was combined with the BICEP2/Keck likelihood~\citep{bk18:2021} in order to provide constraints on the tensor-to-scalar ratio $r$.

\begin{figure}[htbp!]
    \centering
    \includegraphics[width=.9\columnwidth]{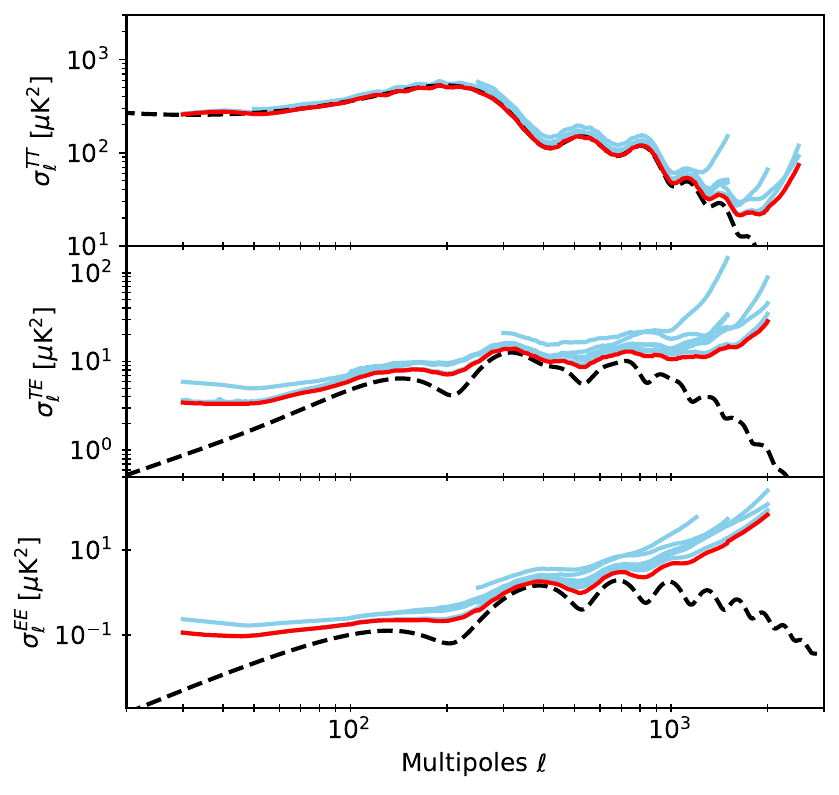}
    \caption{Uncertainties on each angular cross-power spectrum (blue lines) and their combination (red line) for the \planck $TT$ (\textit{top}), $TE$ (\textit{middle}), and $EE$ (\textit{bottom}) data, compared to sample variance for 75\,\% of the sky (black dashed line).}
    \label{fig:sigl}
\end{figure}

For the \highl\ power-spectrum analysis, \hillipop, we use a multi-frequency Gaussian likelihood approximation using sky maps at three frequencies (100, 143 and 217\GHz), while the channel at 353\GHz\ is used to derive a template for the dust power spectrum contaminating the CMB signal at large scales.
\hillipop is one of the likelihoods developed within the \planck collaboration and used to analyse previous \planck data sets \citep{planck2013-p08,planck2014-a13}. 
Here, we describe a new version adapted to PR4 and called ``\hillipop~V4.2.'' It differs from the previous one essentially by using a larger sky fraction (covering 75\,\% of the sky) and a refined model for the foregrounds (in particular for point sources and dust emission).  We specifically use High-Frequency Instrument ``detsets'', which are splits of the detectors at each frequency into specific subsets.
We compute cross-spectra for each of the CMB modes ($TT$, $TE$, $EE$), cross-correlating the two detset maps at each of the three \planck channels dominated by the CMB (100, 143, and 217\GHz), together with their associated covariance. As illustrated in Fig.~\ref{fig:sigl}, the variance of the cross-spectra is close to the expected sample variance for 75\,\% of the sky in temperature for $TT$, while the impact of the \planck noise in polarization is more visible in $TE$ and $EE$. However, at those scales ($\ell$ < 2000), \planck PR4 is the most sensitive data set for CMB anisotropies as of today.

The cross-spectra are then co-added into cross-frequency spectra and compared through a Gaussian likelihood to a model taking into account Galactic as well as extragalactic residual emission on top of the CMB signal. As opposed to other \planck\ likelihoods, \hillipop\ considers all cross-frequency power spectra. Even if the \planck PR4 data set is dominated by CMB anisotropies over the entire range of multipoles considered in the \highl\ likelihood ($30\,{<}\,\ell\,{<}\,2500$), using all cross-frequency spectra allows us to check the robustness of the results with respect to our knowledge of the astrophysical foregrounds. Indeed, even if the basic \lcdm parameters are insignificantly affected by the details of the foreground modelling, the constraints on extensions to \lcdm might depend more critically on the accuracy of the foreground description. Moreover, future ground-based experiments, measuring smaller scales than those accessible by \planck, will be even more sensitive to extragalactic foregrounds.

We begin this paper by summarizing the \planck PR4 pipeline (\npipe), focusing on the improvements as compared to PR3 (Sect.~\ref{sec:data}). Then, in Sect.~\ref{sec:spectra}, we explain how the angular power spectra are calculated, and describe the masks we use, the multipole ranges, the pseudo-$C_\ell$ algorithm, and the covariance matrix. The \lollipop likelihood is briefly described in Sect.~\ref{sec:lowl}, with reference to \citet{tristram:2021,tristram:2022}. The \hillipop likelihood is described in Sect.~\ref{sec:hillipop}, including details of foreground modelling and instrumental effects. Results on the parameters for the \lcdm model are described and commented on in Sect.~\ref{sec:lcdm}. Constraints on foreground parameters and instrumental parameters are discussed in Sects.~\ref{sec:foregrounds} and \ref{sec:nuisances}, respectively. Section~\ref{sec:consistency} is dedicated to consistency checks with respect to previous \planck results. Finally we explore some extensions to \lcdm in Sect.~\ref{sec:ext}, specifically the lensing consistency parameter \Alens, the curvature \Omegak, the effective number of neutrino species \Neff, and the sum of neutrino masses \mnu.

\section{The \planck PR4 data set}
\label{sec:data}

The \planck sky measurements used in this analysis are the PR4 maps available from the \planck Legacy Archive\footnote{\myurl{pla.esac.esa.int}} (PLA) and from the National Energy Research Scientific Computing Center (NERSC).\footnote{\myurl{portal.nersc.gov/project/cmb/planck2020}} They have been produced with the \npipe processing pipeline, which creates calibrated frequency maps in temperature and polarization from both the \planck Low-Frequency Instrument (LFI) and the High-Frequency Instrument (HFI) data. As described in \citet{planck2020-LVII}, \npipe processing includes data from the repointing periods that were neglected in previous data releases. There were additionally several improvements, resulting in lower levels of noise and systematics in both frequency and component-separated maps at essentially all angular scales, as well as notably improved internal consistency between the various frequencies.
Moreover, PR4 also provides a set of ``End-to-End'' Monte Carlo simulations processed with \npipe, which enables the characterization of potential biases and the uncertainties associated with the pipeline.

To compute unbiased estimates of the angular power spectra, we perform cross-correlations of two independent splits of the data. As shown in \citet{planck2020-LVII}, the most appropriate split for the \planck data is represented by the detset maps, comprising two subsets of maps with nearly independent noise characteristics, made by combining half of the detectors at each frequency. This was obtained by processing each split independently, in contrast to the split maps produced in the previous \planck releases. We note that time-split maps (made from, e.g., ``odd $-$ even rings'' or ``half-mission data'') share the same instrumental detectors, and therefore exhibit noise correlations due to identical spectral bandpasses and optical responses. As a consequence, the use of time-split maps gives rise to systematic biases in the cross-power spectra \citep[see section~3.3.3 in][]{planck2016-l05}, as well as underestimation of the noise levels in computing the half-differences (which needed to be compensated by a rescaling of the noise in PR3, as described in appendix~A.7 of \citealt{planck2016-l03}). For this reason, we cross-correlate using detset splits only.

\begin{figure}[htbp!]
\centering
\includegraphics[width=\columnwidth]{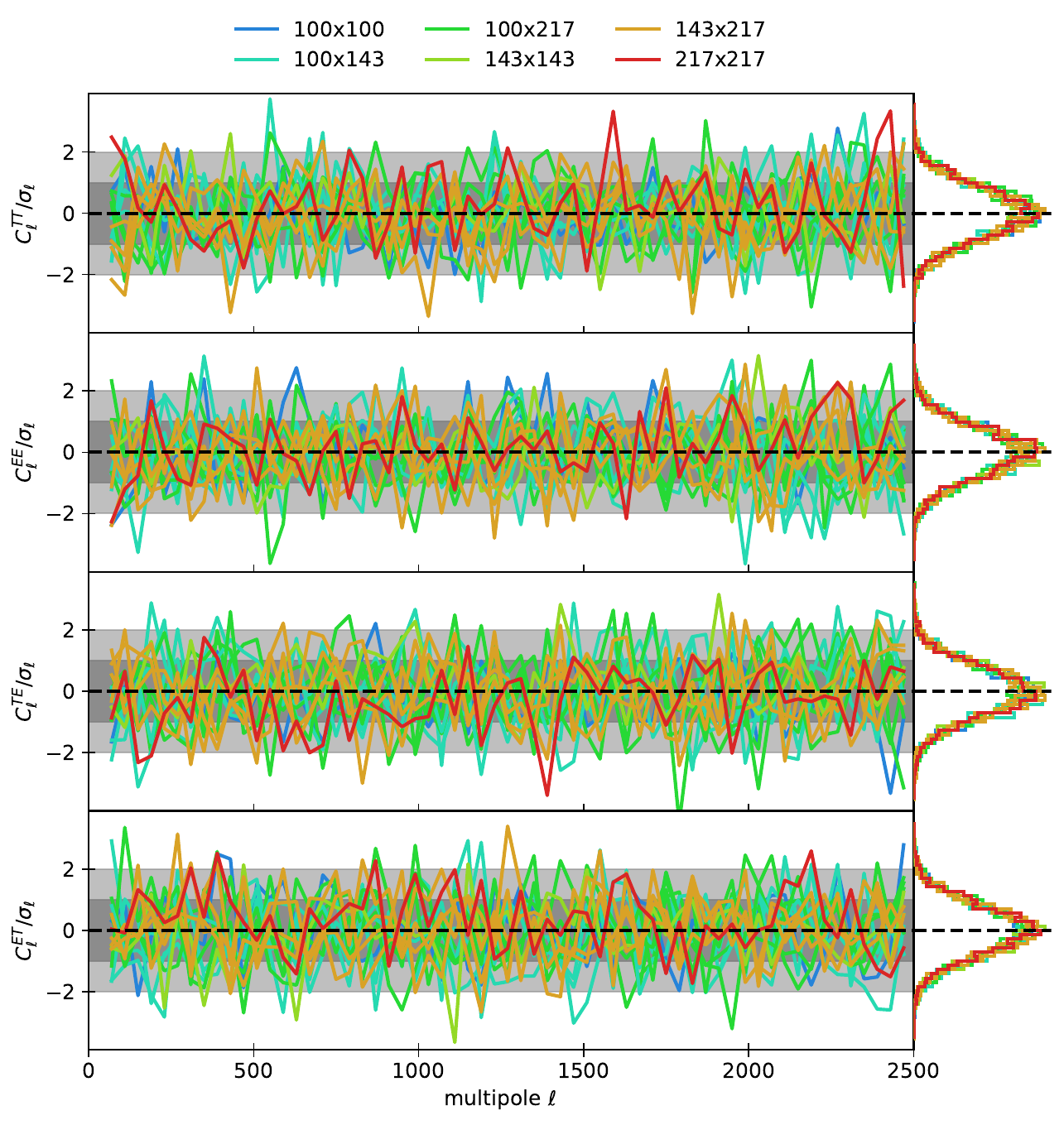}
\caption{Detset cross-spectra for half-ring differences computed on 75\,\% of the sky, divided by their uncertainties. From top to bottom we show $TT$, $EE$, $TE$, and $ET$. Spectra are binned with $\Delta\ell=40$. The projections on the right show the distribution for each unbinned spectrum over the range $\ell=30$--2500.}
\label{fig:hrd}
\end{figure}

Nevertheless, in order to verify the level of noise correlation between detsets, we computed the detset cross-power spectra from the half-ring difference maps, which we show in Fig.~\ref{fig:hrd}. 
The spectra are computed on 75\,\% of the sky and are fully compatible with zero, ensuring that any correlated noise is much smaller than the uncorrelated noise over the range of multipoles from $\ell=30$ to 2500. As discussed above, this test is not sensitive to correlations at scales smaller than the half-ring period. Indeed, if both halves of a ring are affected by the same systematic effect, it will vanish in the half-ring difference map and thus will not be tested in cross-correlation with another detset. 

\section{\planck PR4 angular power spectra}
\label{sec:spectra}

\subsection{Large-scale polarized power spectra}
Foregrounds are stronger in polarization relative to the CMB than in temperature, and cleaning the \planck frequencies using $C_\ell$ templates in the likelihood (as done at small scales) is not accurate enough, especially at large angular scales. In order to clean sky maps of polarized foregrounds, we use the \commander component-separation code \citep{eriksen2008}, with a model that includes three polarized components, namely the CMB, synchrotron emission, and thermal dust emission. \commander is run on each detset map independently, as well as on each realization from the PR4 Monte Carlo simulations.

We then compute unbiased estimates of the angular power spectra by cross-correlating the two detset-cleaned maps. 
We compute power spectra using an extension of the quadratic maximum-likelihood estimator \citep{tegmark:2001} adapted for cross-spectra in \citet{vanneste:2018}. At multipoles below 40, this has been shown to produce unbiased polarized power spectra with almost optimal errors. We use downgraded $\nside\,{=}\,16$ maps \citep{gorski2005} after convolution with a cosine apodizing kernel $b_\ell = \frac{1}{2}\left\{1+\cos\pi(\ell-1)/(3\nside-1)\right\}$.
The signal is then corrected with the PR4 transfer function, to compensate for the filtering induced by the degeneracies between the signal and the templates for systematics used in the mapmaking procedure~\citep[see][]{planck2020-LVII}.

\begin{figure}[!htbp]
    \center
    \includegraphics[width=0.9\columnwidth]{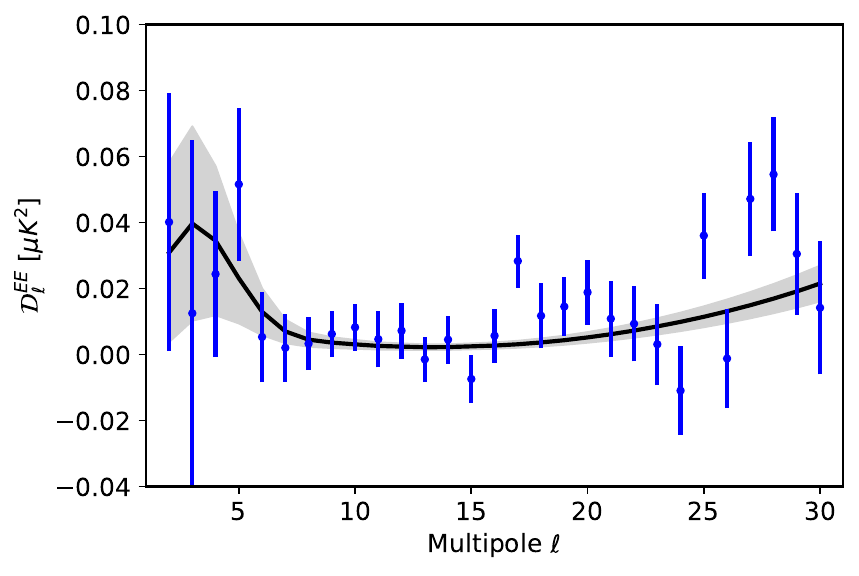}
    \caption{$EE$ power spectrum of the CMB computed on 50\,\% of the sky with the PR4 maps at low multipoles \citep{tristram:2021}. The \planck 2018 \lcdm model is plotted in black. The grey band represents the associated sample variance. Error bars are deduced from the PR4 Monte Carlo simulations.}
    \label{fig:xqml}
\end{figure}

The resulting power spectrum estimated on the cleanest 50\,\% of the sky is plotted in Fig.~\ref{fig:xqml} up to $\ell\,{=}\,30$ \citep[for more details, see][]{tristram:2021}. We also performed the same estimation on each of the PR4 simulations and derive the $\ell$-by-$\ell$ covariance matrix that is then used to propagate uncertainties in \lollipop, the \lowl\ CMB likelihood described in Sect.~\ref{sec:lowl}.

\subsection{Small-scale power spectra}

\subsubsection{Sky fractions}
\label{sec:masks}

For small scales ($\ell\,{>}\,30$), we are using detset maps at frequencies 100, 143, and 217\GHz, and we select only a fraction of the sky in order to reduce the contamination from Galactic foregrounds. 
The main difference with respect to the masks used for the previous versions of \hillipop \citep[][]{couchot_te:2017} lies in two points: the new Galactic masks allow for a larger sky fraction; and the point-source mask is common to all three frequencies. 
The resulting masks applied to each frequency are made of a combination of four main components, which we now describe.

\paragraph{Galactic mask.}
We apply a mask to remove the region of strongest Galactic emission, adapted to each frequency. We can keep a larger sky fraction at the lowest frequency (100\GHz) where the emission from the Galactic sources is low. Since \planck uncertainty is dominated by sample variance up to multipole $\ell\,{\simeq}\,1800$ in temperature (and $\ell\,{\simeq}\,1100$ in $TE$ polarization), this allows us to reduce the sampling variance by ensuring a larger sky fraction. However, we remove a larger fraction of the sky for the highest frequency channel (217\GHz) since it is significantly more contaminated by Galactic dust emission.

We build Galactic masks using the \planck 353-GHz map as a tracer of the thermal dust emission in intensity. In practice, we smooth the \planck 353-GHz map to increase the signal-to-noise ratio before applying a threshold that depends on the frequency.  Masks are then apodized using a $1\pdeg0$ Gaussian taper for power spectra estimation. For polarization, \planck dust maps show that the diffuse emission is strongly related to the Galactic magnetic field at large scales \citep{planck2014-XIX}. However, at the smaller scales that matter here ($\ell\,{>}\, 30$), the orientation of dust grains is driven by local turbulent magnetic fields that produce a polarization intensity approximately proportional to the total intensity dust map. We thus use the same Galactic mask for polarization as for temperature.

\paragraph{CO mask.}
We apply a mask for CO line emission. We consider the combination of maps of the two lines in the \planck frequency bands at 115 and 230\GHz.
We smooth the \planck reconstructed CO maps to 30~arcmin before applying a threshold at $2\,{\rm K}\,{\rm km}\,{\rm s}^{-1}$. The resulting masks are then apodized at 15\,arcmin.
The CO masks remove 17\,\% and 19\,\% of the sky at 100 and 217\GHz, respectively, although the removed pixels largely fall within the Galactic masks.

\paragraph{Point-sources mask.}
We use a common mask for the three CMB frequencies to cover strong sources (both radio and infrared).
In contrast to the masks used in \plik or \camspec, the point-source mask used in our analysis relies on a more refined procedure that preserves Galactic compact structures and ensures the completeness level at each frequency, but with a higher flux cut on sources (approximately 340, 250, and 200\,mJy at 100, 143, and 217\GHz, respectively). The consequence is that these masks leave a slightly greater number of unmasked extragalactic sources, but more accurately preserve the power spectra of dust emission (see~Sect.~\ref{sec:lik:model}). 
We apodize these masks with a Gaussian taper of 15\,arcmin.
We produce a single point-source mask as the combination of the three frequency masks; in total, this removes 8.3\,\% of the sky.

\paragraph{Large objects.}
We mask a limited number of resolved objects in the sky, essentially nearby galaxies including the LMC, SMC, and M31, as well as the Coma cluster. This removes less than 0.4\,\% of the sky.
\\

We use the same mask for temperature and polarization. Even though masking point sources in polarization is not mandatory (given the \planck noise in $EE$, and $TE$); this makes the computation of the covariance matrix much simpler while not removing a significant part of the sky.

\begin{figure}[htbp!]
    \centering
    \includegraphics[width=\columnwidth]{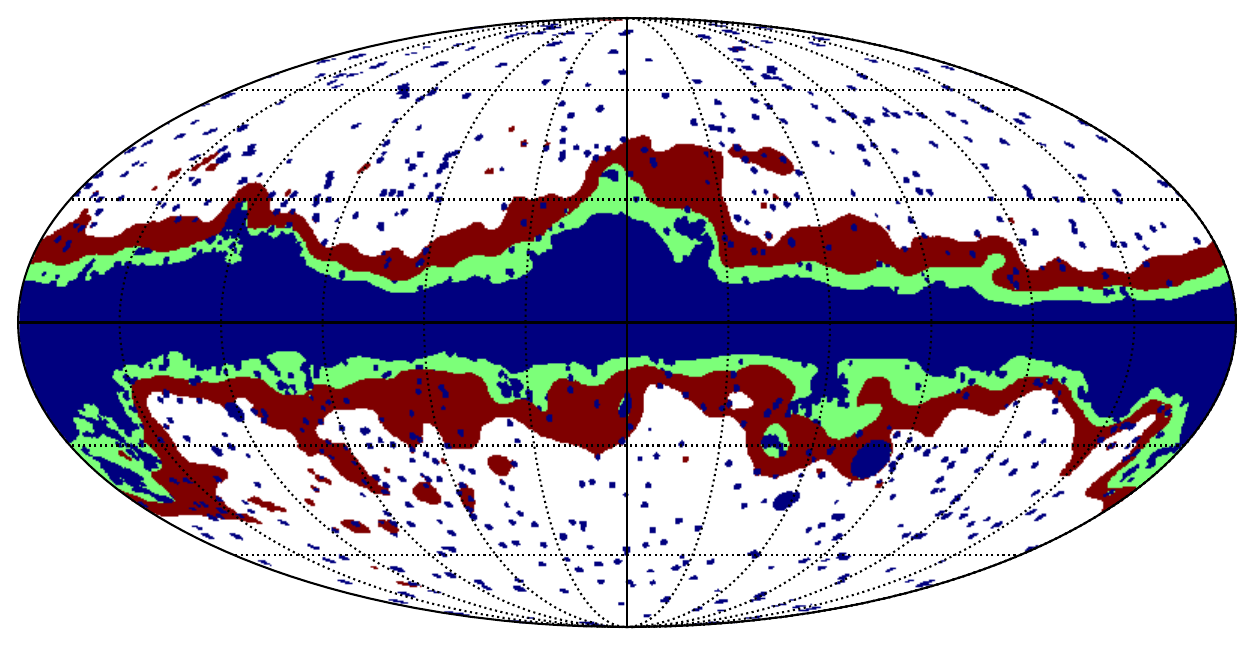}
    \caption{Sky masks used for \hillipop V4.2 as a combination of a Galactic mask (blue, green, and red for the 100, 143, and 217\GHz\ channel, respectively), a CO mask, a point-source mask, and a mask removing nearby galaxies. The effective sky fractions remaining at 100, 143 and 217\GHz\ are 75\,\%, 66\,\%, and 52\,\%, respectively.}
    \label{fig:masks}
\end{figure}

The Galactic masks ultimately used for \hillipop V4.2 cover 20\,\%, 30\,\%, and 45\,\% of the sky for the 100, 143 and 217\GHz\ channels, respectively. After combining with the other masks, the effective sky fraction used for computing cross-spectra are 75\,\%, 66\,\%, and 52\,\%, respectively (see Fig.~\ref{fig:masks}).
The sky fractions retained for the likelihood analysis are about 5\,\% larger than the ones used in the previous version of \hillipop.
Before extending the sky fraction used in the likelihood, we have checked the robustness of the results and the goodness-of-fit (through estimating $\chi^2$) using various combinations of Galactic masks (see Sect.~\ref{sec:consistency}).

\subsubsection{PR4 small-scale spectra}
\label{sec:xpol}

\begin{figure*}[htbp!]
    \center
    \includegraphics[width=0.65\textwidth]{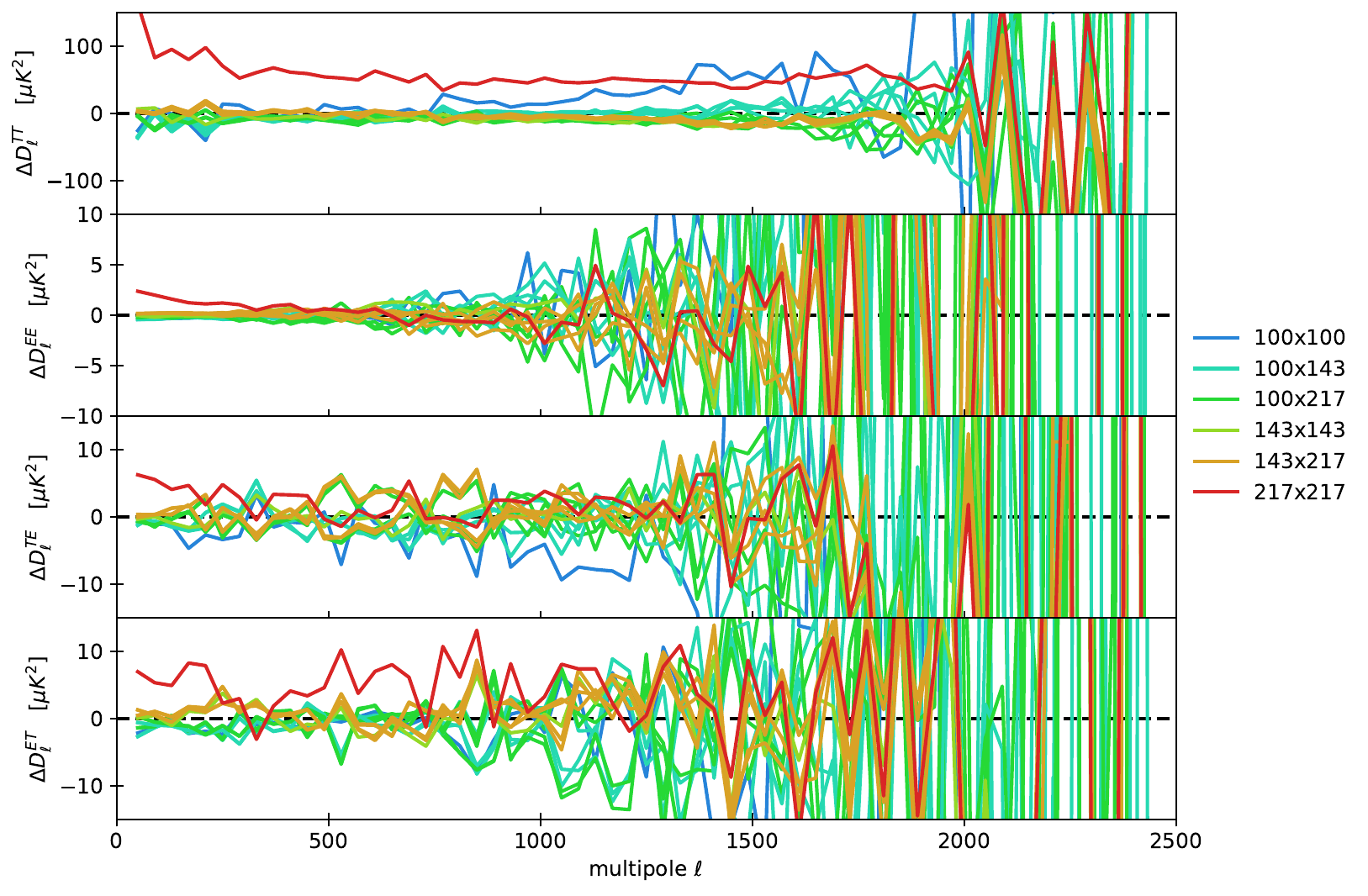}
    \caption{Frequency cross-power spectra with respect to the mean spectra for $TT$, $EE$, $TE$, and $ET$. Spectra are binned with $\Delta\ell=40$ for this figure.}
    \label{fig:cl_diff}
\end{figure*}

We use \xpol \citep[an extension to polarization of \texttt{Xspect}, described in][]{tristram:2005} to compute the cross-power spectra in temperature and polarization ($TT$, $EE$, and $TE$). \xpol is a pseudo-$C_{\ell}$ method \citep[see e.g.,][]{hivon:2002,brown:2005} that also computes an analytical approximation of the $C_\ell$ covariance matrix directly from data.\footnote{\myurl{gitlab.in2p3.fr/tristram/Xpol}} 
Using the six maps presented in Sect.~\ref{sec:data}, we derive the 15 cross-power spectra for each CMB mode, as outlined below: one each for 100$\times$100, 143$\times$143, and 217$\times$217; and four each for 100$\times$143, 100$\times$217, and 143$\times$217.

From the coefficients of the spherical harmonic decomposition of the ($I$,$Q$,$U$) masked maps $\vec{\tilde a}_{\ell m}^X = \{\tilde a^T_{\ell m},\tilde a^E_{\ell m},\tilde a^B_{\ell m}\}$, we form the pseudo cross-power spectra between map $i$ and map $j$,
\begin{equation}
    \widetilde{\vec C}_\ell^{ij} = \frac{1}{2\ell+1} \sum_{m} \vec{\tilde a}^{i*}_{\ell m} \vec{\tilde a}^{j}_{\ell m} \, ,
\end{equation}
where the vector $\vec{\widetilde C}_\ell$ includes the four modes $\{\widetilde C^{\,TT}_\ell,\widetilde C^{\,EE}_\ell, \widetilde C^{\,TE}_\ell,\widetilde C^{\,ET}_\ell\}$.
We note that the $TE$ and $ET$ cross-power spectra do not carry the same information, since computing $T$ from  map $i$ and $E$ from map $j$ is different from computing $E$ from map $j$ and $T$ from $i$. They are computed independently and averaged afterwards using their relative weights for each cross-frequency.
The pseudo-spectra are then corrected for beam and sky fraction using a mode-mixing coupling matrix, $\tens{M}$, which depends on the masks used for each set of maps \citep{peebles:1973,hivon:2002},
\begin{equation}
    \widetilde{\vec C}_\ell^{ij} = (2\ell'+1) \tens{M}^{ij}_{\ell \ell'} \vec{C}^{ij}_{\ell'} \, .
    \label{eq:master}
\end{equation}

The \planck data set suffers from leakage of $T$ to $E$ and $B$, essentially due to beam mismatch between the detectors used to construct the ($I,Q,U$) maps. We debias the beam leakage together with the beam transfer function using the beam window functions evaluated with \quickpol \citep{hivon:2017}. We use the \quickpol\ transfer functions specifically evaluated for PR4, since data cuts, glitch flagging, and detector noise weights all differ from earlier \Planck\ releases. Once corrected, the cross-spectra are inverse-variance averaged for each frequency pair in order to form six unbiased (though correlated) estimates of the angular power spectrum.

The resulting cross-frequency spectra are plotted in Fig.~\ref{fig:cl_diff} with respect to the $C_\ell$ average. For $TT$, the agreement between the different spectra is better than $20\muK^2$, except (as expected) for the 100$\times$100 and the 217$\times$217 cases, which are affected by residuals from point sources and Galactic emission (for the latter). In $EE$, only the 217$\times$217 case is affected by Galactic emission residuals at low multipoles, but the spectra are still consistent at the few $\mu{\rm K}^2$ level. For $TE$ and $ET$, we can see various features at the level of $10\muK^2$ (especially for the $100T\times100E$ and $217E\times217T$ spectra).
Even though the consistency between the cross-frequencies is very good, the likelihood presented in Sect.~\ref{sec:hillipop} will take into account those residuals from foreground emission.

\subsubsection{Multipole ranges}
\label{sec:multipoles}
The \hillipop likelihood covers the multipoles starting from $\ell_{\rm min}=30$ up to $\ell_{\rm max}=2500$ in temperature and $\ell_{\rm max}=2000$ in polarization. The multipoles below $\ell<30$ are considered in the \lowl\ likelihoods (\commander and \lollipop, see Sect.~\ref{sec:lowl}).

Table~\ref{tab:multipoles} gives the \hillipop multipole ranges, $[\ell_{\rm min},\ell_{\rm max}]$, considered for each of the six cross-frequencies in $TT$, $TE$, and $EE$. The multipole ranges used in the likelihood analysis have been chosen to limit the contamination by Galactic dust emission at low $\ell$ and instrumental noise at high $\ell$. 
In practice, we ignore the lowest multipoles for cross-spectra involving the 217\GHz\ map, where dust contamination is the highest, and cut out multipoles higher than $\ell = 1500$ for cross-spectra involving the 100\GHz\ channel given its high noise level.

In total, the number of multipoles considered is now 29\,758 for $TT+TE+EE$, to be compared to the number in the \hillipop analysis of PR3, which was 25\,597. 
The spectra are sample-variance limited up to $\ell \simeq 1800$ in $TT$ and $\ell \simeq 1100$ in $TE$, while the $EE$ mode is essentially limited by instrumental noise.

\begin{table}[!htbp]
    \begin{center}
    \begin{tabular}{lccc}
        \hline
        \hline
        \noalign{\vskip 1pt}
Channels & $TT$ & $TE$ & $EE$\\
\hline
\noalign{\vskip 1pt}
100$\times$100 & [30,1500] & [30,1500] & [100,1200]\\
100$\times$143 & [30,1500] & [30,1500] & [30,1500]\\
100$\times$217 & [250,1500] & [100,1500] & [250,1500]\\
143$\times$143 & [50,2000] & [30,2000] & [30,2000]\\
143$\times$217 & [250,2500] & [200,2000] & [250,2000]\\
217$\times$217 & [250,2500] & [300,2000] & [250,2000]\\
\hline
\noalign{\vskip 1pt}
 & 10646 & 9816 & 9296\\
        \hline
    \end{tabular}
    \caption{Multipole ranges used in the \hillipop analysis and corresponding number of $\ell$s available ($n_{\ell}=\ell_{\rm max}-\ell_{\rm min}+1$). The total number of $\ell$s across all spectra is $29\,758$.}
    \label{tab:multipoles}
    \end{center}
\end{table}

\subsubsection{The covariance matrix}
\label{sec:covmat}
We use a semi-analytical estimate of the $C_\ell$ covariance matrix computed using \xpol. The matrix captures the $\ell$-by-$\ell$ correlations between all the power spectra involved in the analysis. The computation relies directly on data for the estimates. It follows that contributions from noise (correlated and uncorrelated), sky emission (from astrophysical and cosmological origin), and the sample variance are implicitly taken into account in this computation without relying on any model or simulations.

The covariance matrix $\tens\Sigma$ of the cross-power spectra is directly related to the covariance $\tens{\widetilde\Sigma}$ of the pseudo cross-power spectra through the coupling matrices:
\begin{eqnarray}
    \tens{\Sigma}_{\ell_1\ell_2}^{ab,cd} 
    \equiv \left<\Delta C_{\ell}^{ab}\Delta C_{\ell^\prime}^{cd*}\right>
    = \left(M_{\ell\ell_1}^{ab}\right)^{-1} \widetilde{\tens{\Sigma}}_{\ell_1\ell_2}^{ab,cd} \left(M_{\ell^\prime\ell_2}^{cd*}\right)^{-1},
\end{eqnarray}
with $(a,b,c,d) \in \{T,E\}$ for each map.

The matrix $\tens{\widetilde\Sigma}$, which gives the correlations between the pseudo cross-power spectra ($ab$) and ($cd$), is an N-by-N matrix (where $N=4\ell_{\rm max}$) and reads
\begin{eqnarray*}
\label{eq:correlation}
    \tens{\widetilde\Sigma}_{\ell\ell^\prime}^{ab,cd} &\equiv& \left<\Delta\tilde{C}_{\ell}^{ab}\Delta\tilde{C}_{\ell^\prime}^{cd*}\right>
    =  \left<\tilde{C}_{\ell}^{ab}\tilde{C}_{\ell^\prime}^{cd*}\right>-\tilde{C}_{\ell}^{ab}\tilde{C}_{\ell^\prime}^{cd*}   \\
    &=& \sum_{mm^\prime} \frac{
        \left<\tilde{a}_{\ell m}^{a}\tilde{a}_{\ell^\prime m^\prime}^{c*}\right>\left<\tilde{a}_{\ell m}^{b*}\tilde{a}_{\ell^\prime m^\prime}^{d}\right>+
        \left<\tilde{a}_{\ell m}^{a}\tilde{a}_{\ell^\prime m^\prime}^{d*}\right>\left<\tilde{a}_{\ell m}^{b*}\tilde{a}_{\ell^\prime m^\prime}^{c}\right>
    }{(2\ell+1)(2\ell^\prime+1)}  ,
\end{eqnarray*}
by expanding the four-point Gaussian correlation using Isserlis' formula (or Wick's theorem).
We compute $\tens{\widetilde\Sigma}$ for each pseudo cross-spectra block independently, which includes $\ell$-by-$\ell$ correlation and four spectral mode correlations $\{TT,EE,TE,ET\}$.

Each two-point correlation of pseudo-$\alm$s can be expressed as the convolution of $\vec C_\ell$ with a kernel that depends on the polarization mode considered:
\begin{align*}
    \VEV{ \tilde a^{T_a*}_{\lm}\tilde a^{T_b}_{\lmprime}} &= \sum_{\lmun} \Cl{\ell_1}{T_aT_b} \W{0}{T_a}{\lm}{\lmun} \W{0}{T_b*}{\lmprime}{\lmun},
    \\
    \VEV{ \tilde a^{E_a*}_{\lm} \tilde a^{E_b}_{\lmprime}} &= \frac{1}{4} \sum_\lmun
    \left\{
        \Cl{\ell_1}{E_aE_b} \W{+}{E_a*}{\lm}{\lmun} \W{+}{E_b}{\lmprime}{\lmun}
        + \Cl{\ell_1}{B_aB_b} \W{-}{E_a*}{\lm}{\lmun} \W{-}{E_b}{\lmprime}{\lmun}
    \right\},
    \\
    \VEV{\tilde a^{T_a*}_{\lm} \tilde a^{E_b}_{\lmprime}} &= \frac{1}{2} \sum_\lmun
        \Cl{\ell_1}{T_a E_b} \W{0}{T_a*}{\lm}{\lmun} \W{+}{E_b}{\lmprime}{\lmun},
\end{align*}
where the kernels $W^{0}$, $W^{+}$, and $W^{-}$ are defined as linear combinations of products of $Y_{\ell m}$ of spin 0 and $\pm 2$, weighted by the spherical transform of the window function in the pixel domain (the apodized mask). As suggested in \citet{efstathiou:2006}, by neglecting the gradients of the window function and applying the completeness relation for spherical harmonics \citep{varshalovich:1988}, we can reduce the products of four $W$s into kernels similar to the coupling matrix $\tens{M}$ defined in Eq.~(\ref{eq:master}).
In the end, the blocks of the $\tens{\Sigma}$ matrices are
\begin{eqnarray*}
    \tens{\Sigma}^{T_aT_b,T_cT_d}
    &\simeq
    \Cl{\ell\ell'}{T_aT_c}\Cl{\ell\ell'}{T_bT_d} \tens{M}_{TT,TT} &+\ \Cl{\ell\ell'}{T_aT_d}\Cl{\ell\ell'}{T_bT_c} \tens{M}_{TT,TT},
    \\
    \tens{\Sigma}^{E_aE_b,E_cE_d}
    &\simeq
    \Cl{\ell\ell'}{E_aE_c}\Cl{\ell\ell'}{E_bE_d} \tens{M}_{EE,EE} &+\ \Cl{\ell\ell'}{E_aE_d}\Cl{\ell\ell'}{E_bE_c} \tens{M}_{EE,EE},
    \\
    \tens{\Sigma}^{T_aE_b,T_cE_d}
    &\simeq
    \Cl{\ell\ell'}{T_aT_c}\Cl{\ell\ell'}{E_bE_d} \tens{M}_{TE,TE} &+\ \Cl{\ell\ell'}{T_aE_d}\Cl{\ell\ell'}{E_bT_c} \tens{M}_{TT,TT},
    \\
    \tens{\Sigma}^{T_aT_b,T_cE_d}
    &\simeq
    \Cl{\ell\ell'}{T_aT_c}\Cl{\ell\ell'}{T_bE_d} \tens{M}_{TT,TT} &+\ \Cl{\ell\ell'}{T_aE_d}\Cl{\ell\ell'}{T_bT_c} \tens{M}_{TT,TT},
    \\
    \tens{\Sigma}^{T_aT_b,E_cE_d}
    &\simeq
    \Cl{\ell\ell'}{T_aE_c}\Cl{\ell\ell'}{T_bE_d} \tens{M}_{TT,TT} &+\ \Cl{\ell\ell'}{T_aE_d}\Cl{\ell\ell'}{T_bE_c} \tens{M}_{TT,TT},
    \\
    \tens{\Sigma}^{E_aE_b,T_cE_d}
    &\simeq
    \Cl{\ell\ell'}{E_aT_c}\Cl{\ell\ell'}{E_bE_d} \tens{M}_{TE,TE} &+\ \Cl{\ell\ell'}{E_aE_d}\Cl{\ell\ell'}{E_bT_c} \tens{M}_{TE,TE},
\end{eqnarray*}
which are thus directly related to the measured auto- and cross-power spectra \citep[see the appendix in][]{couchot_te:2017}. In practice, to avoid any correlation between $C_\ell$ estimates and their covariance, we use a smoothed version of each measured power spectrum (using a Gaussian filter with $\sigma_\ell=5$) to estimate the covariance matrix. 

\begin{figure}[!htbp]
    \centering
    \includegraphics[width=0.95\columnwidth]{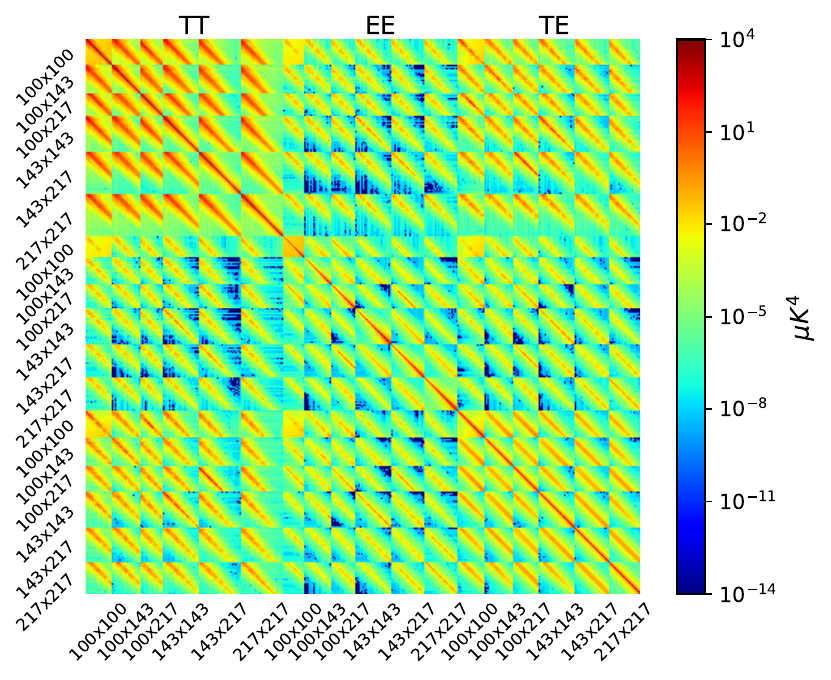}
    \caption{ Full \hillipop covariance matrix, including all correlations in multipoles between cross-frequencies and power spectra. }
    \label{fig:CovMat}
\end{figure}

We finally average the cross-power spectra covariance matrix to form the full cross-frequency power-spectra matrices for the three modes $\{TT,TE,EE\}$. The resulting covariance matrix (Fig.~\ref{fig:CovMat}) has $29\,758\times29\,758$ elements, and is symmetric as well as positive definite. 

This semi-analytical estimation has been tested against Monte Carlo simulations. In particular, we tested how accurate the approximations are in the case of a non-ideal Gaussian signal (due to the presence of small foregrounds residuals), \planck's realistic (low) level of pixel-pixel correlated noise, and the apodization length used for the mask.
We found no deviation to the sample covariance estimated from the 1000 realizations of the full focal plane \planck simulations that include anisotropic correlated noise and foreground residuals. To go further and check the detailed impact from the sky mask (including the choice of the apodization length), we simulated CMB maps from the \planck best-fit $\Lambda$CDM angular power spectrum, to which we added realistic anisotropic Gaussian noise (non-white, but without correlation) corresponding to each of the six data set maps. We then computed their cross-power spectra using the same foreground masks as for the data. A total of $15\,000$ sets of cross-power spectra were produced. 
When comparing the diagonal of the covariance matrix from the analytical estimation with the corresponding simulated variance, a precision better than a few percent is found \citep[see][]{couchot_te:2017}.
Since we are using a Gaussian approximation of the likelihood, the uncertainty of the covariance matrix will not bias the estimation of the  cosmological parameters. The percent-level precision obtained here will then only propagate into a sub-percent error on the variance of the recovered cosmological parameters.

\section{Large-scale CMB likelihoods: \lollipop and \commander}
\label{sec:lowl}

\lollipop (LOw-$\ell$ LIkelihood on POlarized Power spectra) is a \planck\ \lowl\ polarization likelihood based on cross-spectra. It was first applied to \planck PR3 $EE$ data for investigating the reionization history in \citet{planck2014-a25}. It was then upgraded to PR4 data and was described in detail in \citet{tristram:2021} and \cite{tristram:2022}, where it was used to derive constraints on the tensor-to-scalar ratio. 
\lollipop can include $EE$, $BB$, and $EB$ cross-power spectra calculated on component-separated CMB detset maps processed by \commander from the PR4 frequency maps. Here we are focusing only on the $E$-mode component. 

Systematic effects are considerably reduced in cross-correlation compared to auto-correlation, and \lollipop is based on cross-power spectra for which the bias is zero when the noise is uncorrelated between maps. It uses the approximation presented in \citet{hamimeche:2008}, modified as described in \citet{mangilli:2015} to apply to cross-power spectra. The idea is to apply a change of variable $C_\ell \rightarrow X_\ell$ so that the new variable $X_\ell$ is nearly Gaussian-distributed.  Similarly to \citet{hamimeche:2008}, we define
\begin{equation}
  X_\ell = \sqrt{ C_\ell^{\rm f} + O_\ell} \,\, g{\left(\frac{\widetilde{C}_\ell + O_\ell}{C_\ell + O_\ell}\right)} \,\, \sqrt{ C_\ell^{\rm f} + O_\ell} \,,
\label{eq:xell}
\end{equation}
where $g(x)=\sqrt{2(x-\ln(x)-1)}$, $\widetilde{C}_\ell$ are the measured cross-power spectra, $C_\ell$ are the power spectra of the model to be evaluated, $C_\ell^{\rm f}$ is a fiducial CMB model, and $O_\ell$ are the offsets needed in the case of cross-spectra. 
In the case of auto-power spectra, the offsets $O_\ell$ are given by the noise bias effectively present in the measured power spectra. For cross-power spectra, the noise bias is zero, and we use effective offsets defined from the $C_\ell$ noise variance:
\begin{equation}
	\Delta C_\ell \equiv \sqrt{ \frac{2}{2\ell+1}} O_\ell .
\end{equation}

The distribution of the new variable $X_\ell$ can be approximated as Gaussian, with a covariance given by the covariance of the $C_\ell$s.
The likelihood function of the $C_\ell$ given the data $\widetilde{C}_\ell$ is then 
\begin{equation}
  -2\ln P(C_\ell|\widetilde{C}_\ell)=\sum_{\ell \ell'} X^{\sf T}_\ell \tens{M}^{-1}_{\ell \ell'} X_{\ell'}.
\end{equation}
Uncertainties are incorporated into the $C_\ell$ covariance matrix $\tens{M}_{\ell\ell'}$, which is evaluated after applying the same pipeline (including \commander component-separation and cross-spectrum estimation on each simulation) to the Monte Carlo simulations provided in PR4. 
While foreground emission and the cleaning procedure are kept fixed in the simulations (so that we cannot include uncertainties arising from an imperfect foreground model), the resulting $C_\ell$ covariance consistently includes CMB sample variance, statistical noise, and systematic residuals, as well as uncertainties from the foreground-cleaning procedure, together with the correlations induced by masking. 
We further marginalize the likelihood over the unknown true covariance matrix \citep[as proposed in][]{sellentin16} in order to propagate the uncertainty in the estimation of the covariance matrix caused by a limited number of simulations.

\lollipop is publicly available on GitHub.\footnote{\myurl{github.com/planck-npipe/lollipop}} 
In this work, we consider only the information from $E$ modes, and restrict the multipole range from $\ell=2$ to $\ell=30$.

To cover the low multipoles ($\ell\,{<}\,30$) in temperature, we make use of the \commander $TT$ likelihood.
It is based on a Bayesian posterior sampling that combines astrophysical component separation and likelihood estimation, and employs Gibbs sampling to map out the full joint posterior \citep{eriksen2008}. It was extensively used in previous \planck analyses~\citep{planck2013-p08,planck2014-a13}. For the 2018 analysis, the version which is used in this work, \commander makes use of all \planck frequency channels, with a simplified foreground model including CMB, a unique low-frequency power-law component, thermal dust, and CO line emission~\citep[see][]{planck2016-l05}.

\section{Small-scale CMB likelihood: \hillipop}
\label{sec:hillipop}
This section describes \hillipop (High-$\ell$ Likelihood on Polarized Power spectra), including the models used for the foreground residuals and the instrumental systematic residuals. \hillipop was developed for the \planck 2013 results and then applied to PR3 and PR4 \citep[e.g.,][]{planck2014-a13,couchot_alens:2017,tristram:2021}. Here we focus on the latest version of \hillipop, released as V4.2.\footnote{\myurl{github.com/planck-npipe/hillipop}} 
We make use of the 15 cross-spectra computed from the six detset maps at 100, 143, and 217\GHz\ (see Sect.~\ref{sec:spectra}). From those 15 cross-spectra (one each for 100$\times$100, 143$\times$143, and 217$\times$217; four each for 100$\times$143, 100$\times$217, and 143$\times$217), we derive six cross-frequency spectra after recalibration and co-addition, and these are compared to the model. Using all cross-frequencies allows us to break some degeneracies in the foreground domain. However, because \planck spectra are dominated by sample variance, the six cross-frequency spectra are highly correlated. We use the full semi-analytic covariance matrix that includes $\ell$-by-$\ell$ correlation, and $\{TT,TE,EE\}$ mode correlation as described in Sect.~\ref{sec:covmat}.

\subsection{The likelihood approximation}
On the full-sky, the distribution of auto-spectra is a scaled-$\chi^2$ with $2\ell+1$ degrees of freedom. The distribution of the cross-spectra is slightly different \citep[see appendix~A in][]{mangilli:2015}; however, above $\ell\,{=}\,30$ the number of modes is large enough that we can safely assume that the $\widetilde{C}_\ell$ are Gaussian-distributed.
Consequently, for high multipoles the resulting likelihood can be approximated by a multivariate Gaussian, including correlations between the values of $C_\ell$ arising from the cut-sky, and reads
\begin{equation}
    -2 \ln \mathcal{L} = 
        \sum_{\substack{i \leqslant j \\ i' \leqslant j'}} 
        \sum_{\ell\ell'} 
            \vec{R}_\ell^{ij} \, \left[\tens{\Sigma}^{-1}\right]_{\ell\ell^\prime}^{ij,{i'}{j'}} \, 
            \vec{R}_{\ell^\prime}^{{i'}{j'}}
        + \ln | \tens{\Sigma} |,
    \label{eq:likelihood}
\end{equation}
where $\vec{R}^{ij}_\ell = \vec{\widetilde C}^{ij}_\ell - \vec{C}^{ij}_\ell$ denotes the residual of the estimated cross-power spectrum $\vec{\widetilde C}_\ell$ with respect to the model $\vec{C}_\ell$, which depends on the frequencies $\{i,j\}$ and is described in the next section. The matrix $\tens{\Sigma} = \left< \vec{R} \vec{R}^{\tens{T}} \right>$ is the full covariance matrix that includes the instrumental variance from the data as well as the cosmic variance from the model. The latter is directly proportional to the model so that the matrix $\tens{\Sigma}$ should, in principle, depend on the model. 
In practice, given our current knowledge of the cosmological parameters, the theoretical power spectra typically differ from each other at each $\ell$ by less than they differ from the observed $\widetilde{C}_\ell$, so that we can expand $\tens{\Sigma}$ around a reasonable fiducial model. As described in~\citet{planck2013-p08}, the additional terms in the expansion are small if the fiducial model is accurate and leaving it out entirely does not bias the likelihood. Using a fixed covariance matrix $\tens{\Sigma}$, we can drop the constant term $\ln|\tens{\Sigma}|$ and recover nearly optimal variance~\citep[see][]{carron:2013}.
Within the approximations discussed above, we  expect the likelihood to be $\chi^2$-distributed with a mean equal to the number of degrees of freedom $n_{\rm dof} = n_\ell - n_{\rm p}$ ($n_\ell$ being the number of band powers in the power spectra and $n_{\rm p}$ the number of fitted parameters) and a variance equal to $2n_{\rm dof}$.

\subsection{The model}
\label{sec:lik:model}
We now present the model ($\vec{\hat C}_\ell$) used in the likelihood of Eq.~(\ref{eq:likelihood}). The foreground emission is mitigated by masking the part of the sky with high foreground signal (Sect.~\ref{sec:masks}) and using an appropriate choice of multipole range (Sect.~\ref{sec:multipoles}). However, our likelihood function explicitly takes into account residuals of foreground emission in the power spectra, together with the CMB model and instrumental systematic effects. In practice, we consider the model and the data in the form $D_\ell = \ell(\ell+1)C_\ell/2\pi$.

We include in the foregrounds, for the temperature likelihood, contributions from:
\begin{itemize}
    \item Galactic dust; 
    \item cosmic infrared background (CIB);
    \item thermal (tSZ) and kinetic (kSZ) Sunyaev-Zeldovich components; 
    \item Poisson-distributed point sources from radio and infrared star-forming galaxies;
    \item the correlation between CIB and the tSZ effect (tSZ$\times$CIB).
\end{itemize}
We highlight that this new version of \hillipop, labelled V4.2, now includes a model for two point-source components, namely dusty star-forming galaxies and radio sources. Consequently the term ``CIB'' hereafter refers to the clustered part only. For all components, we take into account the bandpass response using effective frequencies as listed in table~4 of \citet{planck2013-p03d}.
Galactic emission from free-free or synchrotron radiation is supposed to be weak at the frequencies considered here (above 100\GHz). Nevertheless, we implemented a model for such emission and were not able to detect any residuals from Galactic synchrotron or free-free emission. In the following, we therefore neglect these contributions.

\begin{figure}[!htbp]
    \centering
    \includegraphics[width=0.9\columnwidth]{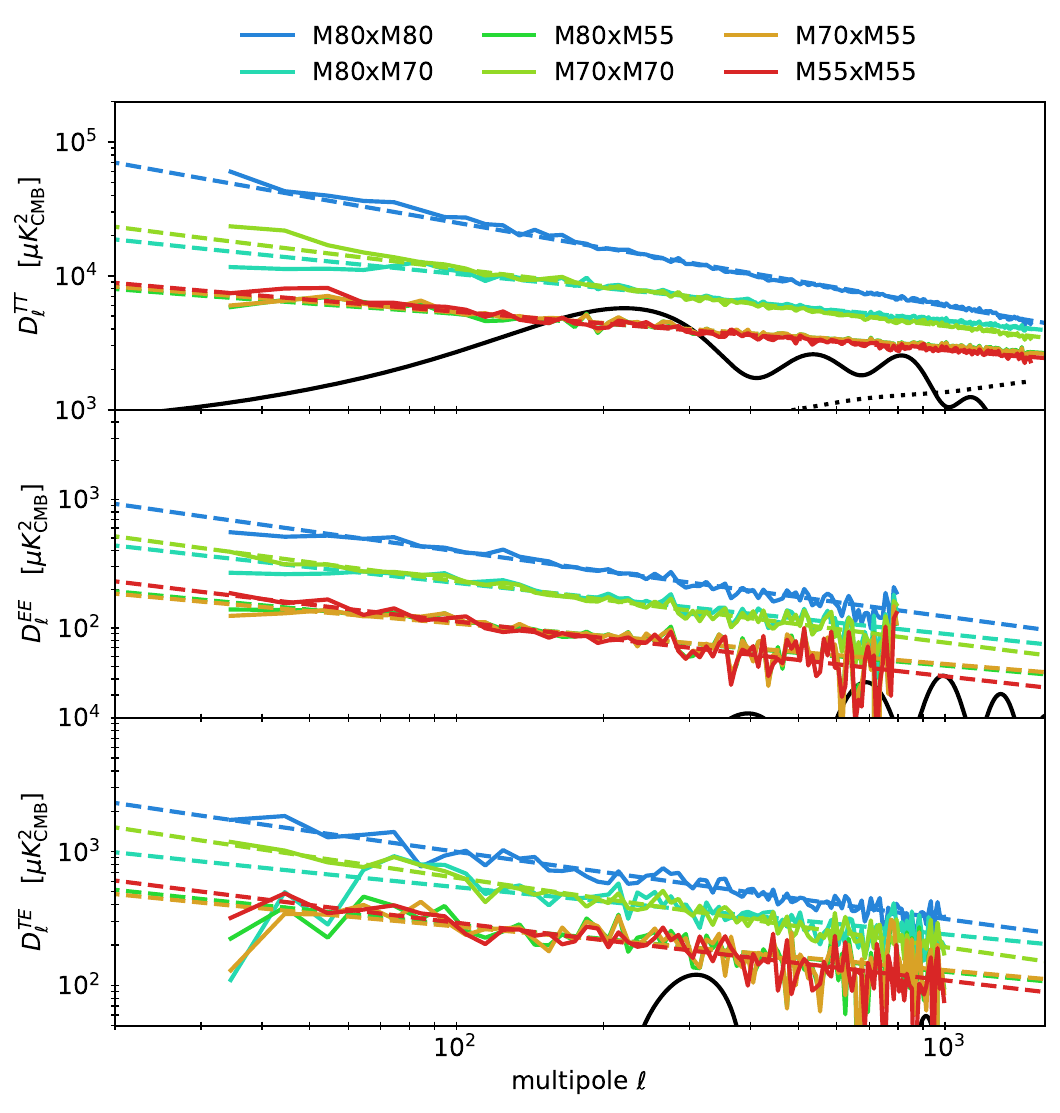}
    \caption{Dust power spectra, $D_\ell = \ell(\ell+1) C_\ell /2\pi$, at 353\GHz\ for $TT$ ({\it top}), $EE$ ({\it middle}), and $TE$ ({\it bottom}). The power spectra are computed from cross-correlation between detset maps at 353\GHz\ for different sets of masks, as defined in Sect.~\ref{sec:masks}, and further corrected for the CMB power spectrum (solid black line) and CIB power spectrum (dashed black line).  The coloured dashed lines are simple fits, as described in the text.}
    \label{fig:dust353}
\end{figure}

\paragraph{Galactic dust emission.}
At frequencies above 100\GHz, Galactic emission is dominated by dust. The dust template is fitted on the \planck 353-GHz data using a power-law model. In practice, we compute the 353-GHz cross-spectra $\vec{\hat C}_\ell^{353Ax353B}$ for each pair of masks $(M_i,M_j)$ associated with the cross-spectra $\nu_i \times \nu_j$ (Fig.~\ref{fig:dust353}). We then subtract the \planck best-fit CMB power spectrum and fit a power-law model with a free constant $A\ell^{\alpha_{\rm d}}+B$ in the range $\ell=[30,1500]$ for $TT$, to account for the unresolved point sources at 353\GHz. A simple power law is used to fit the $EE$ and $TE$ power spectra in the range $\ell=[30,1000]$.
Thanks to the use of the point-source mask (described in Sect.~\ref{sec:masks}), our Galactic dust residual power spectrum is much simpler than in the case of other \planck  likelihoods. Indeed, the point-source masks used in the \planck PR3 analysis removes some Galactic structures and bright cirrus, which induces an artificial knee in the residual dust power spectra around $\ell=200$ \citep[see section~3.3.1 in][]{planck2014-a13}. In contrast, with our point-source mask, the Galactic dust power spectra are fully compatible with power laws (Fig.~\ref{fig:dust353}). While the $EE$ and $TE$ power spectra are directly comparable to those derived in~\citet{planck2014-XXX}, with indices of $\alpha_{\rm d} = -2.3$ and $-2.4$ for $EE$ and $TE$, respectively, the indices for $TT$ vary with the sky fraction considered, ranging from $\alpha_{\rm d} = -2.2$ down to $-2.6$ for the largest sky fraction.

For each polarization mode ($TT$, $EE$, $TE$), we then extrapolate the dust templates at 353\GHz\ for each cross-mask to the cross-frequency considered:
\begin{equation}
    D_\ell^{\rm dust}(\nu\times\nu') = c_\mathrm{\rm dust} \frac{a^{\rm dust}_{\nu}}{a^{\rm dust}_{353}} \frac{a^{\rm dust}_{\nu'}}{a^{\rm dust}_{353}} \mathcal{D}_\ell^{\rm dust}(M_\nu,M_{\nu'}),
\end{equation}
where $a^{\rm dust}_\nu = \nu^{\,\beta_{\rm d}} B_\nu(T_{\rm d})$ is a modified blackbody with $T_{\rm d}$ fixed to $19.6\,$K, while $c_\mathrm{dust}$ and $\beta_{\rm d}$ are sampled independently for temperature and polarization. We use Gaussian priors for the spectral indices $\beta_{\rm d}$ from \citet{planck2014-XXII}, which gives $\beta_{\rm d}^T = \mathcal{N}(1.51,0.01)$ and $\beta_{\rm d}^T = \mathcal{N}(1.59,0.02)$ for temperature and polarization, respectively.
The coefficient $c_\mathrm{dust}$ allows us to propagate the uncertainty from fitting the 353-GHz\ dust spectrum with a power law. We sample $c_\mathrm{dust}$ with a Gaussian prior, $c_\mathrm{dust} = \mathcal{N}(1.0,0.1)$.

\paragraph{Cosmic infrared background (CIB).}
We use a template based on the halo model fitted on \planck and \textit{Herschel} data \citep{planck2013-pip56}, extrapolated with a power-law at high multipoles. The template is rescaled by $A^{\rm CIB}$, the amplitude of the contamination at our reference frequency ($\nu_0=143$\GHz) and $\ell\,{=}\,3000$. The emission law is modelled by a modified blackbody $a_\nu^{\rm CIB} = \nu^{\,\beta_\mathrm{CIB}} B_\nu(T)$ with a fixed temperature ($T=25\,$K) and a variable index $\beta_\mathrm{CIB}$. We use a strong prior $\beta_\mathrm{CIB}=\mathcal{N}(1.75,0.06)$ \citep{planck2013-pip56} and assume perfect correlation between the emission in the frequency range considered (from 100 to 217\GHz),
\begin{equation}
    D_\ell^{\rm CIB}(\nu\times\nu') = A^{\rm CIB} \frac{a_{\nu}^{\rm CIB}}{a_{\nu_0}^{\rm CIB}}\frac{a_{\nu'}^{\rm CIB}}{a_{\nu_0}^{\rm CIB}} \mathcal{D}_\ell^{\rm CIB}.
\end{equation}

\paragraph{Thermal Sunyaev-Zeldovich (tSZ) effect.}
The template for the tSZ emission comes from the halo model fitted on \planck measurements in~\citet{planck2014-a28} and used more recently with PR4 data in \citet{tanimura:2022}. The tSZ signal is parameterized by a single amplitude $A^{\rm tSZ}$, corresponding to the amplitude of the tSZ signal at our reference frequency ($\nu_0=143$\GHz) at $\ell=3000$,
\begin{equation}
    D_\ell^{\rm tSZ}(\nu\times\nu') = A^{\rm tSZ} \frac{a_{\nu}^{\rm tSZ}}{a_{\nu_0}^{\rm tSZ}}\frac{a_{\nu'}^{\rm tSZ}}{a_{\nu_0}^{\rm tSZ}} \mathcal{D}_\ell^{\rm tSZ},
\end{equation}
where $a_{\nu}^{\rm tSZ} = x [e^x+1]/[e^x-1] - 4$ (with $x = h\nu / k_{\rm B} T_{\rm CMB}$).

\paragraph{Kinetic Sunyaev-Zeldovich (kSZ) effect.} 
The kSZ emission is parameterized by $A^{\rm kSZ}$, the amplitude at $\ell\,{=}\,3000$, scaling a fixed template including homogeneous and patchy reionization components from \citet{Shaw12} and \citet{Battaglia13},
\begin{equation}
    D_\ell^{\rm kSZ}(\nu\times\nu') = A^{\rm kSZ} \ \mathcal{D}_\ell^{\rm kSZ} \, .
\end{equation}

\paragraph{Thermal SZ$\times$CIB correlation.} The cross-correlation between the thermal SZ and the CIB is parameterized as
\begin{eqnarray}
    D_\ell^{\rm tSZ\times CIB}(\nu\times\nu')& = & - \xi \sqrt{A^{\rm tSZ} A^{\rm CIB}} \nonumber\\
    && \times\ ( \frac{a_{\nu}^{\rm tSZ}a_{\nu'}^{\rm CIB} + a_{\nu}^{\rm CIB}a_{\nu'}^{\rm tSZ}}{a_{\nu_0}^{\rm tSZ}a_{\nu_0}^{\rm CIB}} )\ \mathcal{D}_\ell^{\rm tSZ\times CIB},
\end{eqnarray}
with $\xi$ the correlation coefficient rescaling the template $\mathcal{D}_\ell^{\rm tSZ\times CIB}$ from \citet{Addison12}.

\paragraph{Point sources.}
Point-source residuals in CMB data sets consist of a combination of the emission coming from radio and infrared sources. For earlier \planck data releases \hillipop used different point-source masks adapted to each frequency. This would require the estimation of the flux cut for each mask in order to use a physical model for the two point-source components. Since the flux-cut estimates are subject to large uncertainties, we used to fit one amplitude for the Poisson term at each cross-frequency in previous \hillipop versions.
In this new version of \hillipop, we adopt a common mask for point sources (see Sect.~\ref{sec:masks}). We then consider a flat Poisson-like power spectrum for each component and use a power law to describe the spectral energy distribution (SED) for the radio sources as $a_{\nu}^{\rm rad} \propto \nu^{-\beta_{\rm s}}$ \citep{tucci:2011} while we use $a_\nu^{\rm IR} = \nu^{\,\beta_{\rm IR}} B_\nu(T)$ \citep{bethermin:2012} for infrared dusty star-forming galaxies. The residual cross-power spectra for point sources are finally
\begin{equation}
C_\ell^{\rm PS}(\nu\times\nu') = A^{\rm rad} \frac{a_\nu^{\rm rad}}{a_{\nu_0}^{\rm rad}} \frac{a_{\nu'}^{\rm rad}}{a_{\nu_0}^{\rm rad}} + A^{\rm IR} \frac{a_\nu^{\rm IR}}{a_{\nu_0}^{\rm IR}} \frac{a_{\nu'}^{\rm IR}}{a_{\nu_0}^{\rm IR}} \, .
\end{equation}

Following \citet{lagache:2020}, radio source emission is dominated at frequencies above about 100\,GHz by radio quasars whose spectral indices can vary from $-1.0$ to $0.0$ \citep{planck2011-6.1,planck2012-VII}. We constrain the SED by fixing $\beta_{\rm s}\,{=}\,-0.8$, following results from \citet{reichardt:2020}. For infrared dusty star-forming galaxies, we adopt $\beta_{\rm IR}$ identical to $\beta_{\rm CIB}$ and $T=25$\,K. The $C_\ell$s are then converted into $D_\ell$s such that the amplitudes $A^{\rm rad}$ and $A^{\rm IR}$ refer to the amplitude of $D_{3000}$ at 143\GHz.
In polarization, we do not include any contribution from point sources, since it is negligible compared to \planck noise for both components \citep{tucci:2004,lagache:2020}.

With the frequencies and the range of multipoles used in the \hillipop likelihood, the foreground residuals are small in amplitude and mostly degenerate in the SED domain. As a result, we choose to set priors on the SED parameters so that the correlation between the amplitudes of residuals is significantly reduced.
The optimization of the foreground model and in particular the determination of the priors adopted for the baseline analysis have been driven by astrophysical knowledge and results from the literature. We have extensively tested the impact of the priors using the \lcdm model as a baseline (without any of its extensions). The results of these tests are discussed in Sect.~\ref{sec:nuisances}.

\subsection{Instrumental effects}
\label{sec:nui}

The main instrumental effects that we propagate to the likelihood are the calibration uncertainties of each of the frequency maps in temperature and polarization (through the polarization efficiency).
As a consequence, we sample five inter-calibration coefficients while fixing as the reference the calibration of the most sensitive map (the first detset at 143\GHz, 143A).
In addition, we sample a \planck calibration parameter $A_{\planck}$ with a strong prior, $A_\planck = \mathcal{N}(1.0000,0.0025)$, in order to propagate the uncertainty coming from the absolute calibration based on the \planck orbital dipole.

We also allow for a recalibration of the polarized maps using polar efficiencies for each of the six maps considered. Those coefficients have been re-estimated in the \npipe processing and we expect them to now be closer to unity and consistent within a frequency channel~\citep{planck2020-LVII}. By default, we fixed the polarization efficiencies to their best-fit values (unity at 100 and 143\GHz\ and 0.975 at 217\GHz, see Sect.~\ref{sec:nuisances} for details).

Angular power spectra have been corrected for beam effects using the beam window functions, including the beam leakage, estimated with \quickpol (see Sect.~\ref{sec:xpol}). With the improvement of the beam-estimation pipeline in \citet{planck2014-a13}, the associated uncertainties have been shown to be negligible in \planck data and are ignored in this analysis.

Discrete sampling of the sky can lead to a small additive (rather than multiplicative) noise contribution known as the ``subpixel'' effect. Its amplitude depends on the temperature gradient within each pixel. With a limited number of detectors per frequency (and even more so per detset), the \planck maps are affected by the subpixel effect. However, estimation of the size of the effect using \quickpol \citep{hivon:2017}, assuming fiducial spectra including CMB and foreground contributions, has shown it to be small \citep{planck2016-l05} and it is therefore neglected in this work.

\section{Results on the 6-parameter \texorpdfstring{$\boldsymbol{\Lambda}$CDM}{LCDM} model}
\label{sec:lcdm}
In this section, we describe the constraints on cosmological parameters in the \lcdm model using the \planck PR4 data. In addition to \hillipop\ (\hlp), we also make use of the \commander low-$\ell$ likelihood~\citep[\lowlT, see][]{planck2016-l04} and the polarized low-$\ell$ $EE$ likelihood \lollipop (\lowlE, discussed in Sect.~\ref{sec:lowl}). We define the following combination of likelihoods for the rest of the paper:
\begin{itemize}
    \item TT, \lowlT+\hlp TT;
    \item TE, \lowlT+\lowlE+\hlp TE;
    \item EE, \lowlE+\hlp EE;
    \item TTTEEE, \lowlT+\lowlE+\hlp TTTEEE.
\end{itemize}
Note that for ``TT'', we only use data from temperature and combine \lowlT+\hlp TT; this is in contrast to \citet{planck2016-l06} and \citet{rosenberg:2022}, in which \lowl\ data from $EE$ are systematically added in order to constrain the reionization optical depth.

The model for the CMB is computed by numerically solving the background and perturbation equations for a specific cosmological model using \texttt{CAMB}~\citep{lewis:2000,howlett:2012}.\footnote{One can equally well use \texttt{CLASS}~\citep{blas:2011} instead, except that the definition of $\theta_*$ differs slightly between the two codes.} In this paper, we consider a \lcdm model with six free parameters describing: the current physical densities of baryons ($\Omega_{\rm b} h^2$) and cold dark matter ($\Omega_{\rm c} h^2$); the angular acoustic scale ($\theta_*$); the reionization optical depth ($\tau$); and the amplitude and spectral index of the primordial scalar spectrum ($A_{\rm s}$ and $n_{\rm s}$).  Here $h$ is the dimensionless Hubble constant, $h=H_0/(100\,{\rm km}\,{\rm s}^{-1}\,{\rm Mpc}^{-1})$.

In addition, we fit for six inter-calibration parameters, seven foreground residual amplitudes in temperature ($c_\mathrm{dust}^T$, $A_\mathrm{radio}$, $A_\mathrm{IR}$, $A_\mathrm{CIB}$, $A_\mathrm{tSZ}$, $A_\mathrm{kSZ}$, and $\xi_\mathrm{SZ\times CIB}$), plus one in polarization ($c_\mathrm{dust}^P$), and three foreground spectral indices ($\beta_\mathrm{dust}^T$, $\beta_\mathrm{dust}^P$, and $\beta_\mathrm{CIB}$). Foreground and instrumental parameters are listed in Table~\ref{tab:params}, together with their respective priors.

To quantify the agreement between the data and the model, we computed the $\chi^2$ values with respect to the best-fit model for each of the data sets using \texttt{Cobaya} \citep{Torrado:2021} with its adaptive, speed-hierarchy-aware MCMC sampler \citep{Lewis:2002,Lewis:2013}.
The $\chi^2$ values and the number of standard deviation from unity are given in Table~\ref{tab:PLK_chi2}. The goodness-of-fit is better than for previous \planck releases, but we still found a relatively large $\chi^2$ value for \hlp TT (corresponding to about 2.7$\,\sigma$), while the \hlp TE and \hlp EE $\chi^2$ values are compatible with unity, at 1.8$\,\sigma$ and 0.1$\,\sigma$, respectively. For the full combination \hlp TTTEEE, we obtain a $\chi^2 = 30495$ for a data size of 29768, corresponding to a 3.02$\,\sigma$ deviation.
As described in \citet{rosenberg:2022}, where the goodness of fit is also somewhat poor (4.07$\,\sigma$ for TT and 4.46$\,\sigma$ for the TTTEEE), this could be explained by a slight misestimation of the instrumental noise, rather than a bias that could be fit by an improved foreground model or a different cosmology. 
However, we emphasize that the level of this divergence is small, since the recovered reduced-$\chi^2$, $\chi^2/n_{\rm d} = 1.02$, shows that the semi-analytical estimation of the covariance of the data is accurate at the percent level.
The goodness-of-fit values for individual cross-spectra are given in Table~\ref{tab:chi2_spectra}.

\begin{table}[!htbp]
        \begin{center}
        \begin{tabular}{lcccc}
        \hline
        \hline
        \noalign{\vskip 2pt}
        Likelihood     & $\chi^2$     &  $n_{\rm d}$        &  $\chi^2/n_{\rm d}$  &  $\delta \sigma(\chi^2)$ \\
        \noalign{\vskip 1pt}
        \hline
        \noalign{\vskip 1pt}
\hlp EE         & 9289 &  9296 & 1.00 & 0.05 \\
\hlp TE         & 10071 &  9816 & 1.03 & 1.82 \\
\hlp TT         & 11044 & 10646 & 1.04 & 2.73 \\
\hlp TTTEEE     & 30495 & 29758 & 1.02 & 3.02 \\
        \hline
        \end{tabular}
        \caption{$\chi^2$ values compared to the size of the data vector ($n_{\rm d}$) for each of the \planck \hillipop likelihoods. Here $\delta \sigma(\chi^2) = (\chi^2/n_{\rm d}-1)/\sqrt{2/n_{\rm d}}$.}
        \label{tab:PLK_chi2}
        \end{center}
\end{table}

\begin{figure*}[ht!]
	\centering
	\includegraphics[width=0.75\textwidth]{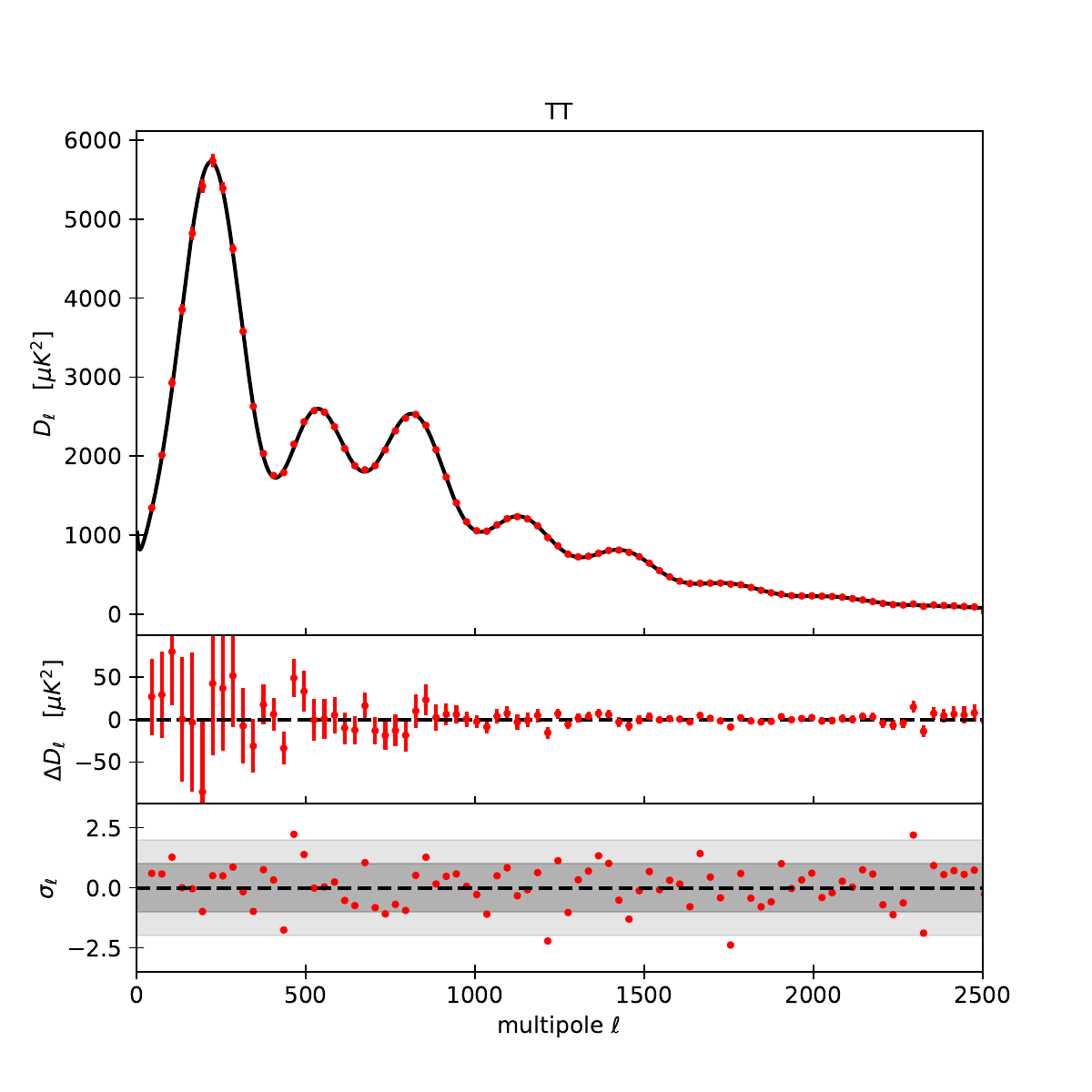}
	\caption{Maximum-likelihood frequency-co-added temperature power spectrum for \hillipop V4.2. For the purposes of this figure, the power spectrum is binned with $\Delta\ell = 30$. The middle panel shows the residuals with respect to the fiducial base-\lcdm cosmology, and the bottom panel shows the residuals normalized by the uncertainties.}
	\label{fig:coadd_TT}

    \includegraphics[width=0.93\columnwidth]{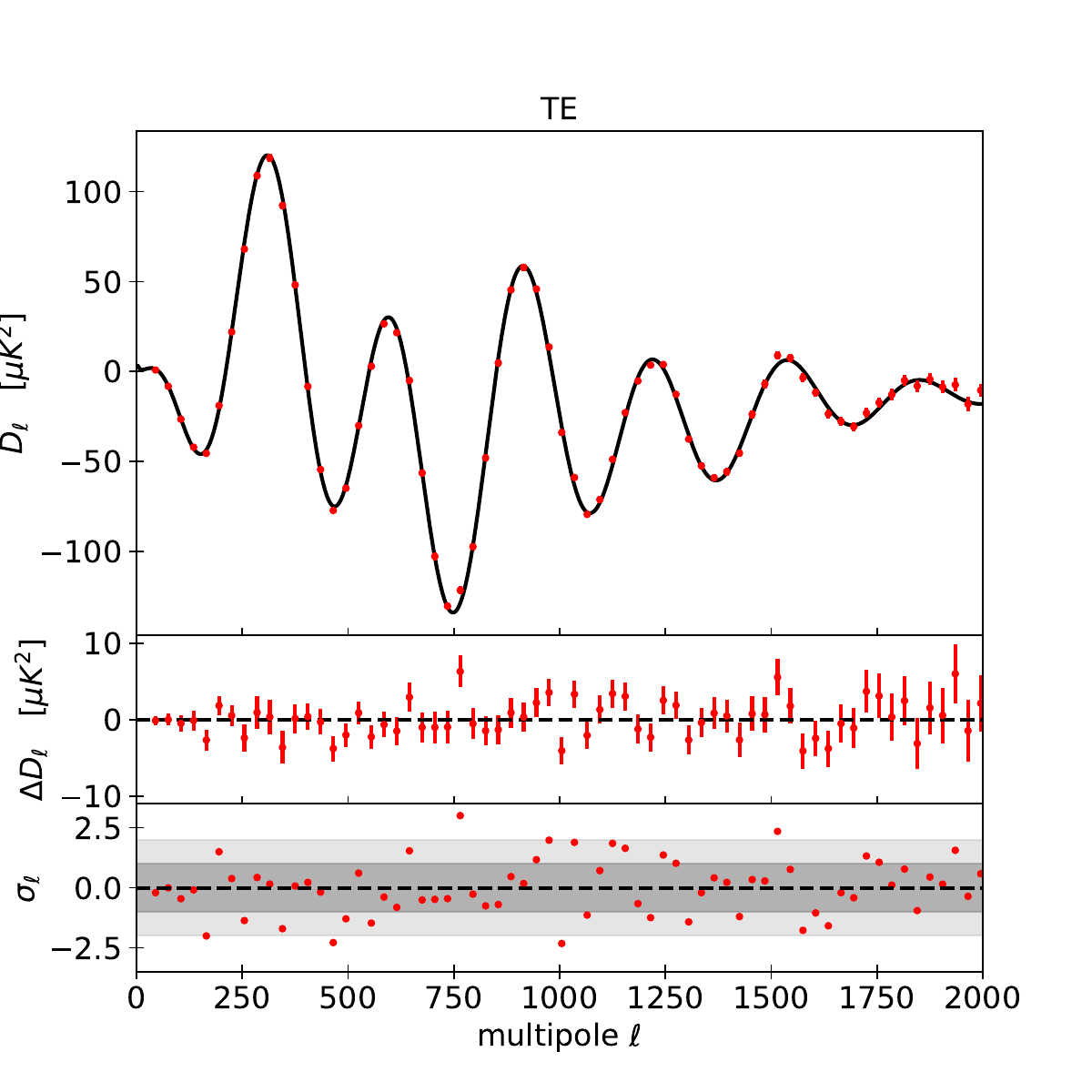}
	\includegraphics[width=0.93\columnwidth]{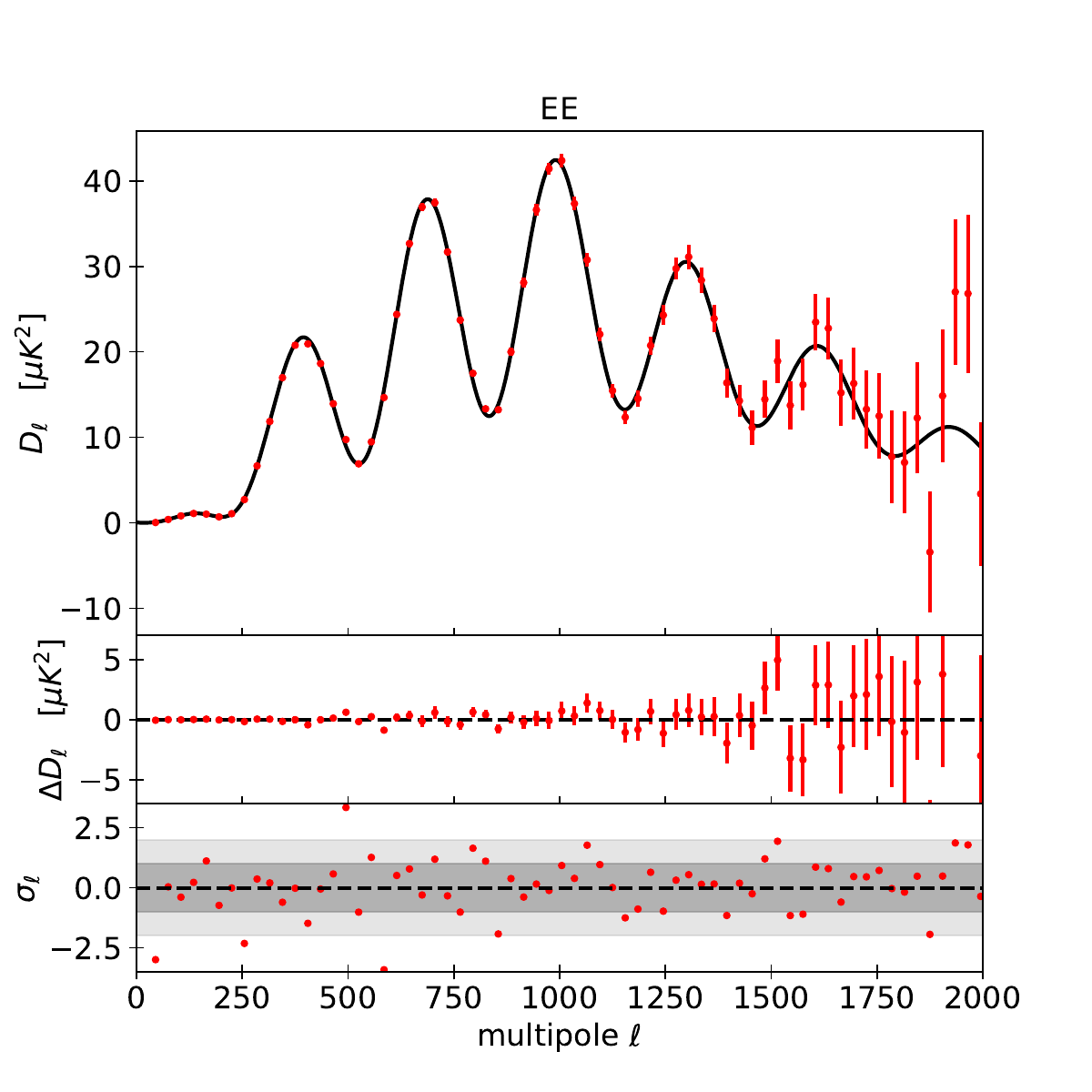}
	\caption{As in Fig.~\ref{fig:coadd_TT}, but for $TE$ (left) and $EE$ (right) power spectra.}
	\label{fig:coadd_TE_EE}
\end{figure*}

Co-added CMB power spectra are shown in Figs.~\ref{fig:coadd_TT} and~\ref{fig:coadd_TE_EE}, for $TT$, $TE$, and $EE$; they are compared to the best-fit obtained with the full TTTEEE combination. \planck spectra are binned with $\Delta\ell=30$ for the plots, but considered $\ell$-by-$\ell$ in the likelihood. The plots also show the residuals relative to the \lcdm best-fit to TTTEEE, as well as the normalized residuals. We cannot identify any deviation from statistical noise or any bias from foreground residuals.

\begin{figure*}[ht!]
	\centering
	\includegraphics[width=.95\textwidth]{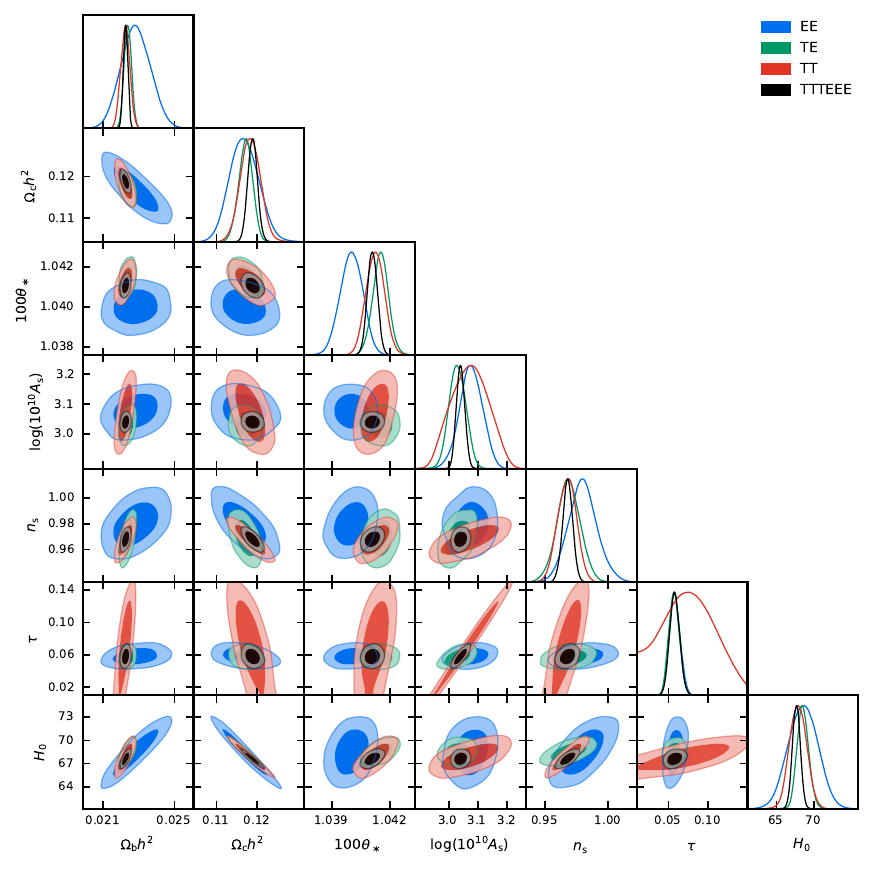}
	\caption{Posterior distributions for the cosmological parameters using power spectra from \planck PR4 with TT (\lowlT+\hlp TT), TE (\lowlT+\lowlE+\hlp TE), EE (\lowlE+\hlp EE), and TTTEEE (\lowlT+\lowlE+\hlp TTTEEE).}
	\label{fig:PLK_lcdm}
\end{figure*}

In Fig.~\ref{fig:PLK_lcdm}, we compare the constraints on \lcdm parameters obtained using TT, TE, and EE and their combination. We find very good consistency between TT and TE, while EE constraints are wider, with a deviation in the acoustic scale~$\theta_*$ toward lower values. This feature of the \planck PR4 data was previously reported in \citet{rosenberg:2022}, in which the authors studied the correlation with other parameters and concluded that this is likely due to parameter degeneracies coupling to residual systematics in EE. 
However, the deviation of $\theta_*$ between EE and TT is now reduced with the increase of the sky fraction enabled by \hillipop V4.2, though still present at the 1.6$\,\sigma$ level.
In addition, we have checked that this shift in $\theta_*$ is not related to any super-sample lensing effect \citep[as described in][]{Manzotti14}, or to any aberration correction \citep[see][]{Jeong14}, both of which are negligible for the large sky fraction considered in the \planck data set.
We note that, interestingly, $\theta_*$ is the only parameter that deviates in EE; the others, including $H_0$, are compatible with TT at much better than 1$\,\sigma$.
Given the weak sensitivity of the \planck $EE$ spectra as compared to $TT$ and $TE$, discrepancies in the EE parameter reconstruction will have little impact on overall cosmological parameter results.

The \hillipop V4.2 constraints on \lcdm cosmological parameters are summarized in Table.~\ref{tab:lcdm}. As compared to the last \planck cosmological results in \citet{planck2016-l06}, the constraints are tighter, with no major shifts. The error bars are reduced by 10 to 20\,\%, depending on the parameter. The reionization optical depth is now constrained at close to the 10\,\% level:
\begin{equation}
    \tau = 0.058 \pm 0.006.
\end{equation}
This is the result of the \npipe treatment of the PR4 data associated with the low-$\ell$ likelihood \lollipop \citep[see][]{planck2020-LVII}.

For the constraint on the Hubble constant, we obtain
\begin{equation}
    H_0 = (67.64 \pm 0.52) \, \text{km\,s$^{-1}$\,Mpc$^{-1}$},
\end{equation}
consistent with previous \planck results and still significantly lower than the local distance-ladder measurements, which typically range from $H_0 = 70$ to $76$, depending on the data set and the calibration used for the first step of the distance  ladder~\citep[see for instance][]{snowmass:2022}.

The amplitude of density fluctuations is
\begin{equation}
    \sigma_8 = 0.8070 \pm 0.0065,
\end{equation}
compatible with PR3 results ($\sigma_8 = 0.8120 \pm 0.0073$) but lower by $0.5\,\sigma$. 
The matter density, $\Omega_\mathrm{m}$ also shifts by roughly $1\,\sigma$, so that
\begin{equation}
    S_8 \equiv \sigma_8 (\Omega_{\rm m}/0.3)^{0.5} = 0.819 \pm 0.014.
\end{equation}
Compared to PR3 ($S_8=0.834 \pm 0.016$), this shift to a lower value of $S_8$ brings it closer to the measurements derived from galaxy clustering and weak lensing from the Dark Energy Survey Year~3 analysis~\citep[$S_8 = 0.782 \pm 0.019$, for \lcdm with fixed $\mnu$,][]{abbott:2022}, decreasing the CMB versus large-scale structure tension on $S_8$ from 2.1$\,\sigma$ to 1.5$\,\sigma$.

\begin{figure}[!htbp]
\centering
\includegraphics[width=0.95\columnwidth]{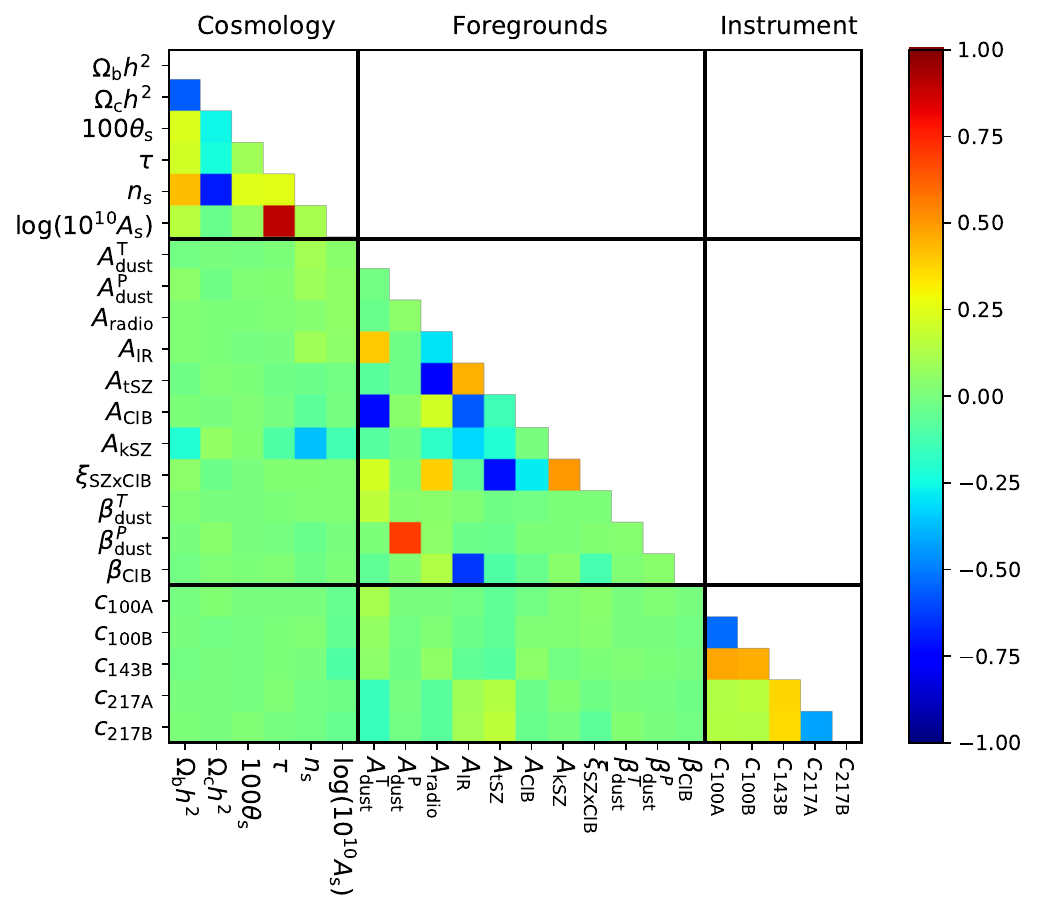}
\caption{Correlation matrix for the fitted parameters of the combined \hillipop likelihood TTTEEE. The first block corresponds to cosmological parameters from the \lcdm model, the second block gathers the foreground parameters, and the last block shows the instrumental parameters.}
\label{fig:lcdm_corr}
\end{figure}

Before discussing results on the foreground parameters (Sect.~\ref{sec:foregrounds}) and instrumental parameters (Sect.~\ref{sec:nuisances}), we show in Fig.~\ref{fig:lcdm_corr} the correlation matrix for the fitted parameters. We can see that foreground parameters are only weakly correlated with the cosmological parameters and the inter-calibrations. This strengthens the robustness of the results with respect to the foreground model and ensures very low impact on cosmology.

\begin{table*}[!htbp]
        \begin{center}
        \begin{tabular}{lcccc}
        \hline
        \hline
        \noalign{\vskip 2pt}
        Parameter     & TT     &  TE & EE & TTTEEE\\
        \hline
        \noalign{\vskip 2pt}
$\Omega_\mathrm{b}h^2$ & $0.02224 \pm 0.00025$ & $0.02238 \pm 0.00020$ & $0.02283 \pm 0.00081$ & $0.02226 \pm 0.00013$ \\
$\Omega_\mathrm{c}h^2$ & $0.1183 \pm 0.0024$ & $0.1172 \pm 0.0017$ & $0.1169 \pm 0.0035$ & $0.1188 \pm 0.0012$ \\
$100\theta_\ast$ & $1.04123 \pm 0.00046$ & $1.04151 \pm 0.00041$ & $1.04001 \pm 0.00059$ & $1.04108 \pm 0.00026$ \\
$\mathrm{log}(10^{10}A_\mathrm{s})$ & $3.073 \pm 0.061$ & $3.030 \pm 0.028$ & $3.077 \pm 0.039$ & $3.040 \pm 0.014$ \\
$n_\mathrm{s}$ & $0.9678 \pm 0.0072$ & $0.9689 \pm 0.0091$ & $0.9797 \pm 0.0110$ & $0.9681 \pm 0.0039$ \\
$\tau$ & $0.0753 \pm 0.0322$ & $0.0572 \pm 0.0065$ & $0.0582 \pm 0.0066$ & $0.0580 \pm 0.0062$ \\
\noalign{\vskip 1pt}
\hline
\noalign{\vskip 2pt}
$H_0$ & $67.89 \pm  1.11$ & $68.49 \pm  0.76$ & $68.49 \pm  1.91$ & $67.64 \pm  0.52$ \\
$\sigma_8$ & $0.8186 \pm 0.0221$ & $0.7973 \pm 0.0129$ & $0.8149 \pm 0.0189$ & $0.8070 \pm 0.0065$ \\
$S_8$ & $0.826 \pm 0.024$ & $0.795 \pm 0.021$ & $0.814 \pm 0.044$ & $0.819 \pm 0.014$ \\
$\Omega_\mathrm{m}$ & $0.3059 \pm 0.0147$ & $0.2983 \pm 0.0099$ & $0.2995 \pm 0.0226$ & $0.3092 \pm 0.0070$ \\
       \noalign{\vskip 1pt}
       \hline
       \noalign{\vskip 1pt}
        \end{tabular}
        \caption{Parameter constraints in the 6-parameter \lcdm model for each data set and their combination, using \hillipop V4.2 in addition to \commander and \lollipop at low $\ell$. We report mean values and symmetrical 68\,\% confidence intervals.}
        \label{tab:lcdm}
        \end{center}
\end{table*}

\section{Foreground parameters}
\label{sec:foregrounds}

\begin{figure}[!htbp]
	\centering
	\includegraphics[width=\columnwidth]{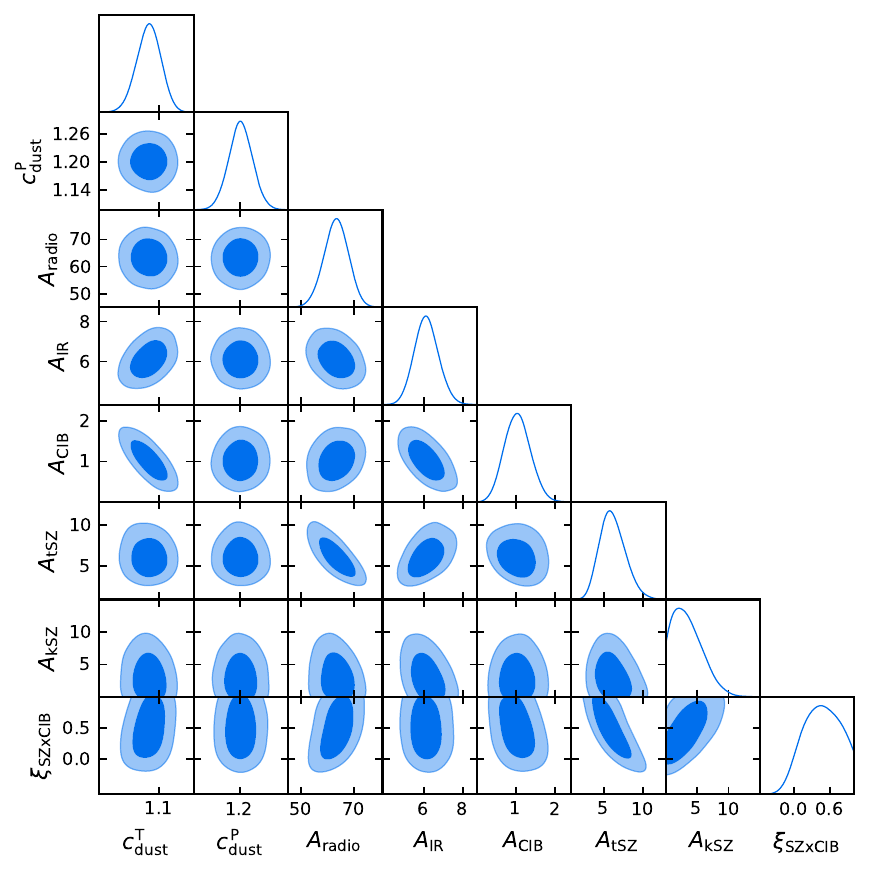}
	\caption{Posteriors for foreground amplitudes. Units are $\mu{\rm K}^2$ normalized at $\ell = 3000$ and $\nu = 143$\GHz.}
	\label{fig:hlp_foregrounds}
\end{figure}

All \planck cross-spectra are dominated by the CMB signal at all the scales we consider. This is illustrated for $TT$ in Fig.~\ref{fig:TT_components} of Appendix~\ref{app:components}, where we show each component of the model fitted in the likelihood with the best-fit parameters for the six cross-frequencies. It is also true for $TE$ and $EE$.
Thanks to the multi-frequency analysis, we are able to break degeneracies related to the fact that some foreground-component power spectra are very similar. 
The resulting marginalized posteriors are plotted in Fig.~\ref{fig:hlp_foregrounds}.
With the choice made for the multipole range and sky fraction, the \planck PR4 data set is sensitive to the CIB, the tSZ, and residual point sources (radio at 100\GHz\ and infrared at 217\GHz). Very low multipoles are sensitive to residuals from Galactic dust emission, especially at 217\GHz.

We detect the emission of radio point sources at better than $16\,\sigma$. The preferred radio power in $D_\ell$ at $\ell=3000$ for 143\GHz\ is
\begin{eqnarray}
    A_\mathrm{radio} &=& (63.3 \pm 4.7)\,\mu {\rm K}^2,
\end{eqnarray}
with a population spectral index for the radio power fixed to $\beta_\mathrm{s} = -0.8$, close to the value recovered by the SPT team~\citep[$\beta_\mathrm{s}=-0.76\pm0.15$,][]{reichardt:2020}.
Allowing $\beta_\mathrm{s}$ to vary in \planck data, gives $\beta_\mathrm{s} = -0.54 \pm 0.08$, with a corresponding increase of the amplitude $A_\mathrm{radio}$.  This also impacts the SZ-CIB cross-correlation amplitude with a significant increase of $\xi$.

We obtain a high-significance detection of CIB anisotropies, with amplitudes at 143\GHz\ and $\ell=3000$ given by
\begin{eqnarray}
    A_\mathrm{CIB} &=& (1.03 \pm  0.34) \, \mu {\rm K}^2,\\
    A_\mathrm{IR} &=& (6.07 \pm  0.63) \, \mu {\rm K}^2,
\end{eqnarray}
for the clustered and Poisson parts, respectively. We note that these amplitudes cannot be directly compared to values in previous works because they strongly depend on the prior used for the $\beta_\mathrm{CIB}$ index for the former and on the flux cut applied by the point-source mask for the latter.

The thermal Sunyaev-Zeldovich effect is also significantly detected, with an amplitude at 143\GHz\ and $\ell=3000$ of
\begin{eqnarray}
    A_\mathrm{tSZ} &=& (5.9 \pm 1.7) \, \mu {\rm K}^2.
\end{eqnarray}
This is close to (but somewhat higher than) what is reported in \citet{reichardt:2020}, with $A_\mathrm{tSZ} = (3.42 \pm 0.54)\,\mu{\rm K}^2$, even though the uncertainties are larger. However, it is more closely comparable with ACTpol results, $A_\mathrm{tSZ} = (5.29 \pm 0.66)\,\mu{\rm K}^2$ \citep{choi:2020}.

We find an upper-limit for the kSZ effect, while the correlation between tSZ and CIB is compatible with zero:
\begin{eqnarray}
	A_\mathrm{kSZ} &<& 7.6 \, \mu {\rm K}^2 \quad\quad \text{(at 95\,\%~CL)};\\
	\xi_\mathrm{SZ\times CIB} &=& 0.46 \pm  0.30.
\end{eqnarray}
We note that those last results are about 10 times less sensitive than the constraints from ground-based CMB measurements, such as those from SPT or ACTpol.

\begin{figure}[!htbp]
	\centering
	\includegraphics[width=0.85\columnwidth]{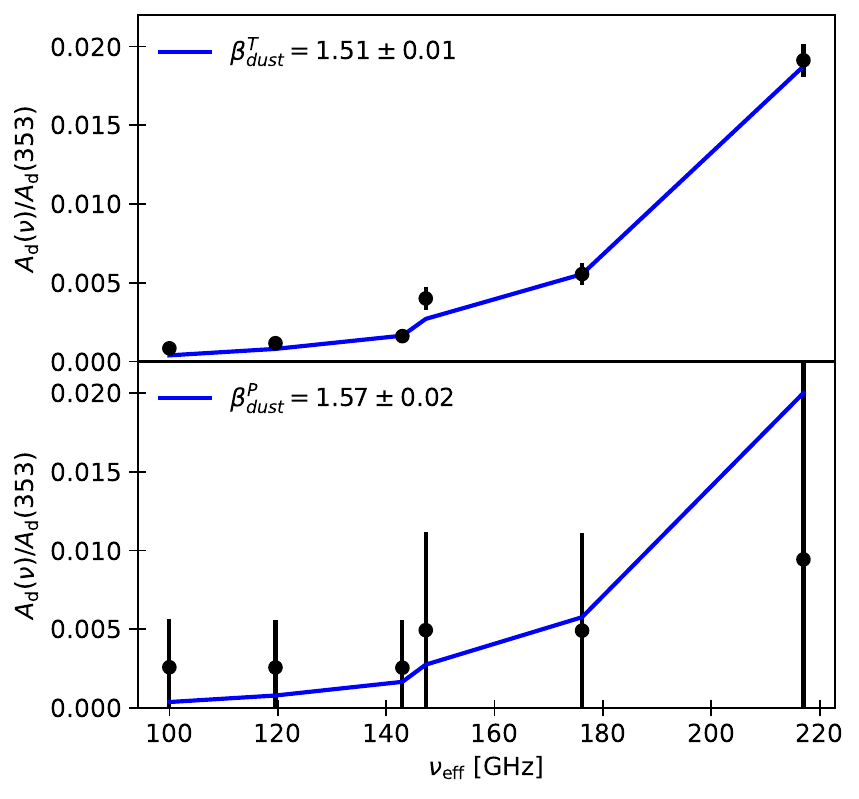}
	\caption{Amplitude of the dust emission relative to 353\GHz\ for a modified-blackbody dust model (blue line) as a function of the effective frequency (computed as the geometric mean of the two frequencies involved), compared to a fit using one amplitude per frequency (black dots). The top panel is for temperature and the bottom panel for polarization.}
	\label{fig:dust_model}
\end{figure}

For the residuals of Galactic dust emission, with priors on the spectral indices driven by \citet{planck2014-XXII}, we find rescaling coefficients $c_\mathrm{dust}$ to be $1.08 \pm 0.03$ and $1.20 \pm 0.03$ for temperature and polarization, respectively. 
This indicates that we recover slightly more dust contamination than our expectations derived from the measurements at 353\GHz, especially in polarization. 
To estimate the impact on the reconstructed parameters (both cosmological and from foregrounds), we sampled the dust amplitudes at each frequency. The constraints are shown in Fig.~\ref{fig:dust_model} for temperature (top) and polarization (bottom). The figure illustrates that we have a good fit of the dust emission in temperature, while we are marginally sensitive to dust residuals in polarization. This explains why, given our prior on the SED for the polarized dust emission, $\beta_{\rm dust}^P = \mathcal{N}(1.59,0.02)$, we recover an amplitude higher than expected.

\begin{figure}[!htbp]
	\centering
	\includegraphics[width=0.85\columnwidth]{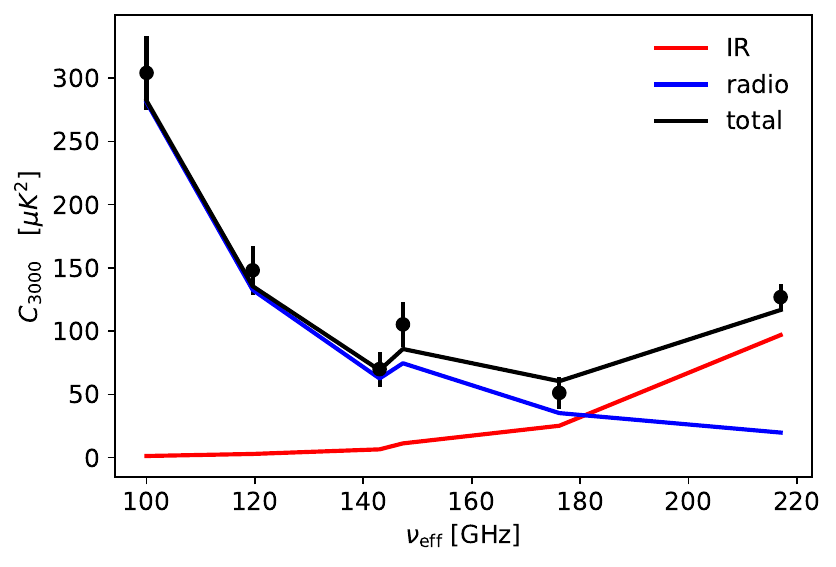}
	\caption{Point-source model as a function of the effective frequency (computed as the geometric mean of the two frequencies involved), compared to the fit of one amplitude per cross-spectrum.}
	\label{fig:ps_model}
\end{figure}

As discussed in Sect.~\ref{sec:lik:model}, \hillipop V4.2 also  includes a 2-component model for point sources.
Figure~\ref{fig:ps_model} shows how the model, as the sum of the two point-source components, matches with the fit with one amplitude for each cross-frequency.

While changing the models as described above, the impact on \lcdm parameters is very limited. We experienced variations of less than 0.11$\,\sigma$ for all \lcdm parameters, with the exception of $n_{\rm s}$, can vary by $0.18\,\sigma$ when changing the model for point sources. Error bars on \lcdm parameters are also stable with respect to foreground modelling, with variations limited to less than 2\,\% (4\,\% for $n_{\rm s}$).

\section{Instrumental parameters}
\label{sec:nuisances}

Inter-calibration parameters are fitted in \hillipop with respect to the first detset at 143\GHz\ (see Sect.~\ref{sec:nui}). The inter-calibrations are recovered at better than the percent level and are compatible with unity. Using the full TTTEEE likelihood, we find
\begin{eqnarray}
c_{\rm 100A} &=& 1.003 \pm 0.007, \\
c_{\rm 100B} &=& 1.004 \pm 0.007, \\
c_{\rm 143B} &=& 1.004 \pm 0.006, \\
c_{\rm 217A} &=& 1.001 \pm 0.008, \\
c_{\rm 217B} &=& 1.001 \pm 0.008.
\end{eqnarray}

\hillipop also allows us to fit for the polarization efficiency even though, by default, those are fixed. Using the full TTTEEE likelihood, we constrain the polarization efficiencies for each map at the percent level. The mean posteriors show polarization efficiencies compatible with unity at better than $1\,\sigma$, except for the two maps at 217\GHz, which differ from unity by about $2\,\sigma$:
\begin{eqnarray}
\eta_{\rm 100A} &=& 0.994 \pm 0.013; \\
\eta_{\rm 100B} &=& 0.987 \pm 0.013; \\
\eta_{\rm 143A} &=& 1.016 \pm 0.013; \\
\eta_{\rm 143B} &=& 1.001 \pm 0.010; \\
\eta_{\rm 217A} &=& 0.978 \pm 0.013; \\
\eta_{\rm 217B} &=& 0.972 \pm 0.014.
\end{eqnarray}
Fixing polarization efficiencies to $1.00$, $1.00$, and $0.975$ (at 100, 143, and 217\GHz, respectively) increases the $\chi^2$ by $\Delta \chi^2 = 36$ for 29758 data points. However, this choice has no effect on either the \lcdm parameters or the foreground parameters.

\section{Consistency between \planck likelihoods}
\label{sec:consistency}
We now investigate the impact of the increased sky fraction used in this new version of \hillipop. We repeat the analysis using more conservative Galactic masks reducing the sky fraction at each frequency by 5\,\% (labelled ``\textit{XL}'') or 10\,\%  (labelled ``\textit{L}'') with respect to our baseline (``\textit{XXL}'', which masks, 20\,\%, 30\,\%, and 45\,\% at 100, 143, and 217\GHz, respectively; see Sect.~\ref{sec:masks} for more details).
Within \lcdm, we obtain similar $\chi^2$ for the fits, demonstrating that the model used in \hillipop V4.2 is valid for the considered sky fraction. For the TTTEEE likelihood, the $\Delta\chi^2$ values are lower than 100 for 29758 data points.

The other \planck likelihood using PR4 data is \camspec and is described in detail in \citet{rosenberg:2022}. Although \camspec is focused on cleaning procedures to build co-added polarization spectra rather than modelling of foreground residuals in cross-frequency spectra, we find consistent constraints at better than the $1\,\sigma$ level. This gives confidence in the robustness of our cosmological constraints. 

\begin{figure}[ht!]
	\centering
	\includegraphics[width=\columnwidth]{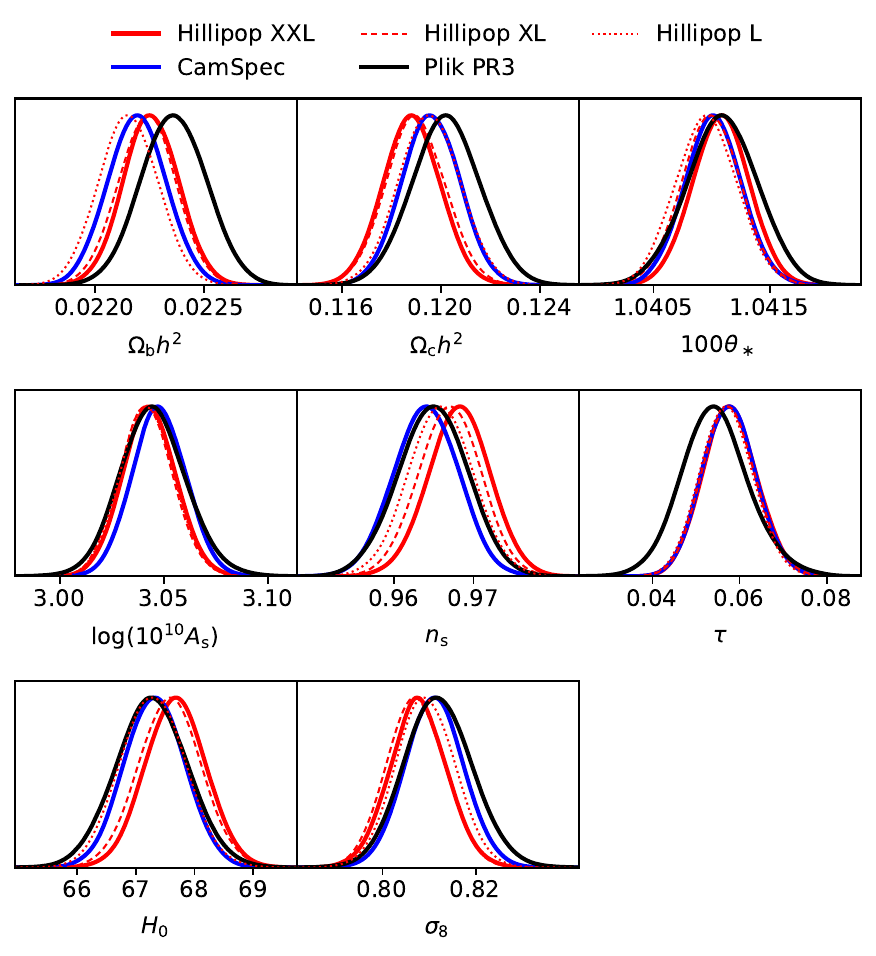}
	\caption{Posterior distributions for the cosmological parameters from PR4 for \hillipop (using different sky fractions labelled {\it L}, {\it XL}, and {\it XXL}) and \camspec, as compared to \planck 2018 (\plik PR3). Likelihoods are considered for the combination of TT+TE+EE, with \lowlT\ and \lowlE\ used at low $\ell$.}
	\label{fig:lcdm_fsky}
\end{figure}

Figure~\ref{fig:lcdm_fsky} shows the 1-d posterior distributions for the \lcdm parameters using different sky fractions. We also compare to the posteriors obtained from \planck PR3 and with \camspec PR4~\citep[where we use \lollipop instead of the polarized low-$\ell$ constraint from PR3 used in][]{rosenberg:2022}.
We find good consistency between the different likelihoods and between the two data sets (PR3 and PR4).

Table~\ref{tab:PR3_PR4} shows the relative difference in the cosmological parameters between \planck 2018~\citep{planck2016-l06} and this work, together with the gain in accuracy. The largest difference with respect to \planck 2018 appears for $\Omega_{\rm c}h^2$, for which \hillipop on PR4 finds a value $1.0\,\sigma$ lower. Associated with \commander and \lollipop, \camspec on PR4 also gives lower $\Omega_{\rm c}h^2$ by $-0.45\,\sigma$. 
The spectral index $n_{\rm s}$ is found to be a bit higher with \hillipop by $0.7\,\sigma$.

As discussed in Sect.~\ref{sec:lcdm}, we obtain a slightly higher value for the Hubble constant ($+0.6\,\sigma$) with $h = 0.6766 \pm 0.0053$, compared to $h = 0.6727 \pm 0.0060$ for PR3. The amplitude of density fluctuations, $\sigma_8$, and the matter density, $\Omega_{\rm m}$, are lower by $0.7\,\sigma$ and $0.8\,\sigma$, respectively, so that $S_8$ is also lower by about $0.9\,\sigma$.
The error bars shrink by more than 10\,\%, with a noticeable gain of 20\,\% for the acoustic scale ($\theta_*$).

\begin{table}[ht!]
        \begin{center}
        \begin{tabular}{lcccc}
        \hline
        \hline
        \noalign{\vskip 1pt}
        Parameter & \phantom{0}$\Delta / \sigma$ & $\Delta \sigma$ \\
        \hline
        \noalign{\vskip 2pt}
$\Omega_\mathrm{b}h^2$ & -0.67 & -13.7\,\% \\
$\Omega_\mathrm{c}h^2$ & -0.99 & -15.2\,\% \\
$100\theta_\ast$ & -0.01 & -16.1\,\% \\
$\mathrm{log}(10^{10}A_\mathrm{s})$ & -0.30 & -12.0\,\% \\
$n_\mathrm{s}$ & +0.75 & -11.0\,\% \\
$\tau$ & +0.44 & -21.4\,\% \\
        \noalign{\vskip 1pt}
\hline
        \noalign{\vskip 2pt}
$H_0$ & +0.61 & -13.7\,\% \\
$\sigma_8$ & -0.70 & -11.5\,\% \\
$S_8$ & -0.89 & -14.2\,\% \\
$\Omega_\mathrm{m}$ & -0.79 & -16.1\,\% \\
       \hline
        \end{tabular}
        \caption{Relative variation and improvement in the error bars between \planck 2018 and this work for each cosmological parameter.}
        \label{tab:PR3_PR4}
        \end{center}
\end{table}

\section{Combination with other data sets}
\label{sec:bao}

We now present some results of our new likelihood in combination with CMB lensing measurements using the \planck\ PR4 data \citep{carron:2022}. We specifically use the conservative range recommended in \citet{carron:2022}, consisting of nine power bins between multipoles of 8 and 400. The addition of the $C_\ell^{\phi\phi}$ information means that we are using all the power spectra available from PR4; hence TTTEEE+lensing provides the best \textit{Planck}-only cosmological constraints currently available.

We supplement this with measurements of the baryon acoustic oscillations (BAOs). This includes data from 6dF~\citep{beutler:2011}, SDSS DR7~\citep[specifically MGS,][]{ross:2015}, and SDSS DR16~\citep[LRG, ELG, QSO, Ly-$\alpha$ auto, and Ly-$\alpha$xQSO,][]{alam:2021}, which also incorporates some constraints on the growth of structures through redshift-space distortions.

\begin{figure}[!htbp]
	\centering
	\includegraphics[width=.9\columnwidth]{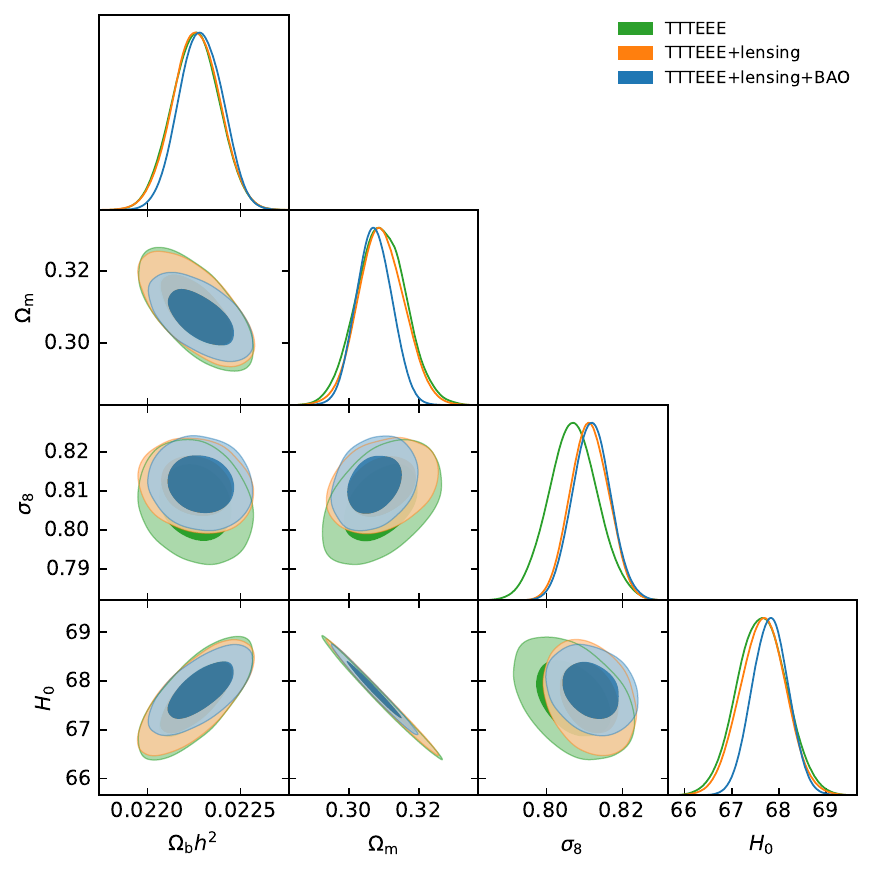}
	\caption{Posterior distributions for some parameters using TTTEEE in combination with lensing and BAO.}
	\label{fig:PR4_BAOlens}
\end{figure}

Table~\ref{tab:lcdm2} presents the constraints on the 6-parameter \lcdm\ model when adding lensing and BAO data.  Figure~\ref{fig:PR4_BAOlens} shows the posterior distribution for the particular subset $\Omega_{\rm b} h^2$, $\Omega_{\rm m}$, $\sigma_8$, and $H_0$.

\begin{table}[!htbp]
\setlength{\tabcolsep}{3pt}
        \begin{center}
        \begin{tabular}{@{\hskip 0pt}l@{\hskip -3pt}ccc@{\hskip -1pt}}
        \hline
        \hline
        \noalign{\vskip 2pt}
        Parameter     & TTTEEE & TTTEEE & TTTEEE    \\
             & &  +lensing & +lensing+BAO    \\
        \hline
        \noalign{\vskip 2pt}
$\Omega_\mathrm{b}h^2$ & $0.02226 \pm 0.00013$ & $0.02226 \pm 0.00013$ & $0.02229 \pm 0.00012$ \\
$\Omega_\mathrm{c}h^2$ & $0.1188 \pm 0.0012$ & $0.1190 \pm 0.0011$ & $0.1186 \pm 0.0009$ \\
$100\theta_\ast$ & $1.04108 \pm 0.00026$ & $1.04107 \pm 0.00025$ & $1.04111 \pm 0.00024$ \\
$\mathrm{log}(10^{10}A_\mathrm{s})$ & $3.040 \pm 0.014$ & $3.045 \pm 0.012$ & $3.048 \pm 0.012$ \\
$n_\mathrm{s}$ & $0.9681 \pm 0.0039$ & $0.9679 \pm 0.0038$ & $0.9690 \pm 0.0035$ \\
$\tau$ & $0.0580 \pm 0.0062$ & $0.0590 \pm 0.0061$ & $0.0605 \pm 0.0059$ \\
\noalign{\vskip 1pt}
\hline
\noalign{\vskip 2pt}
$H_0$ & $67.64 \pm  0.52$ & $67.66 \pm  0.49$ & $67.81 \pm  0.38$ \\
$\sigma_8$ & $0.8070 \pm 0.0065$ & $0.8113 \pm 0.0050$ & $0.8118 \pm 0.0050$ \\
$S_8$ & $0.819 \pm 0.014$ & $0.824 \pm 0.011$ & $0.821 \pm 0.009$ \\
$\Omega_\mathrm{m}$ & $0.3092 \pm 0.0070$ & $0.3092 \pm 0.0066$ & $0.3071 \pm 0.0051$ \\
       \noalign{\vskip 1pt}
       \hline
       \noalign{\vskip 1pt}
        \end{tabular}
        \caption{Parameter constraints in the 6-parameter \lcdm model for each data set and their combination, using \hillipop V4.2 in addition to \commander and \lollipop at low $\ell$, with the addition of CMB lensing and BAO constraints.  We report mean values and symmetrical 68\,\% confidence intervals.}
        \label{tab:lcdm2}
        \end{center}
\end{table}

\section{Extensions}
\label{sec:ext}
We now discuss constraints on some extensions to the base-\lcdm model.

\subsection{Gravitational lensing, \texorpdfstring{$A_\mathrm{L}$}{Alens}}
\label{sec:alens}
We sample the phenomenological extension \Alens\ in order to check the consistency of the \planck PR4 data set with the smoothing of the power spectra by weak gravitational lensing as predicted by the \lcdm model. A mild preference for $\Alens>1$ was seen in the \planck PR1 data \citep{planck2013-p11} and since the analysis of \planck PR2 data \citep{planck2014-a13,planck2014-a15}, \hillipop has provided a significantly lower \Alens\ value than the public \planck likelihood \plik, but still slightly higher than unity. The tension was at the 2.2$\,\sigma$ level for PR3~\citep{couchot_alens:2017}. 
\begin{figure}[ht!]
    \centering
    \includegraphics[width=0.85\columnwidth]{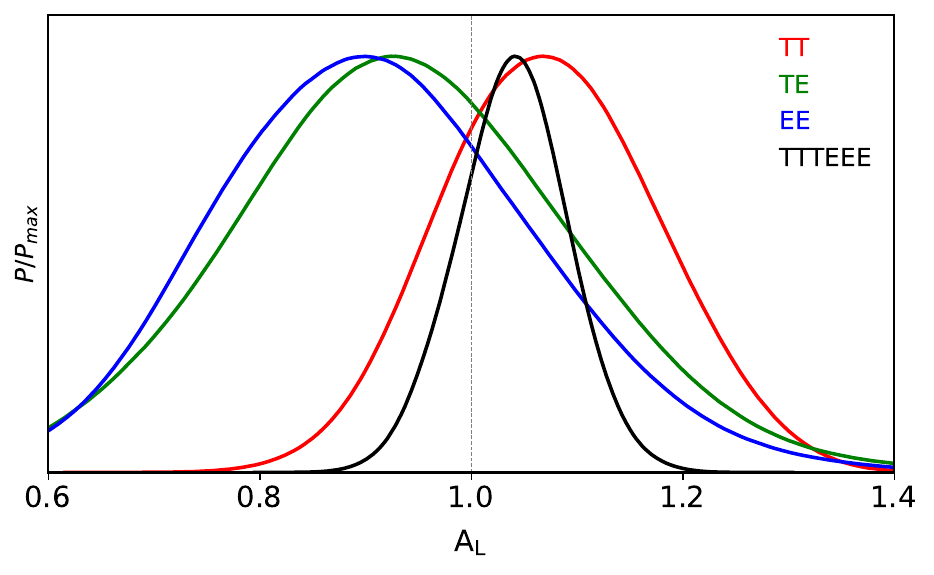}
    \caption{Posterior distributions for \Alens.}
    \label{fig:hlp_alens}
\end{figure}

With \planck PR4, we find results even more compatible with unity compared to previous releases. Indeed for TTTEEE, we now obtain
\begin{equation}
	\Alens = 1.039 \pm 0.052,
\end{equation}
which is compatible with the \lcdm expectation (at the $0.7\,\sigma$ level). As shown in Table~\ref{tab:PR4_alens}, while the results for EE and TE are compatible with unity, the $A_\mathrm{L}$ value for TT is still high by 0.8$\,\sigma$. Figure~\ref{fig:hlp_alens} shows posterior distributions of \Alens\ for each of the mode-spectra and for the TTTEEE combination using \planck PR4.

\begin{table}[ht!]
        \begin{center}
        \begin{tabular}{lcccc}
        \hline
        \hline
        \noalign{\vskip 1pt}
        Likelihood     &  \Alens     &  $\Delta \Alens$    \\
        \hline
        \noalign{\vskip 1pt}
TT & $1.075 \pm 0.102$ & $\phantom{-}0.73\,\sigma$\\
TE & $0.937 \pm 0.158$ & $-0.40\,\sigma$\\
EE & $0.912 \pm 0.150$ & $-0.59\,\sigma$\\
TTTEEE & $1.039 \pm 0.052$ & $\phantom{-}0.75\,\sigma$\\
       \hline
        \end{tabular}
        \caption{Mean values and 68\,\% confidence intervals for \Alens. The significance of the deviation from unity is given in the last column.}
        \label{tab:PR4_alens}
        \end{center}
\end{table}

\begin{figure}[ht!]
    \centering
    \includegraphics[width=0.85\columnwidth]{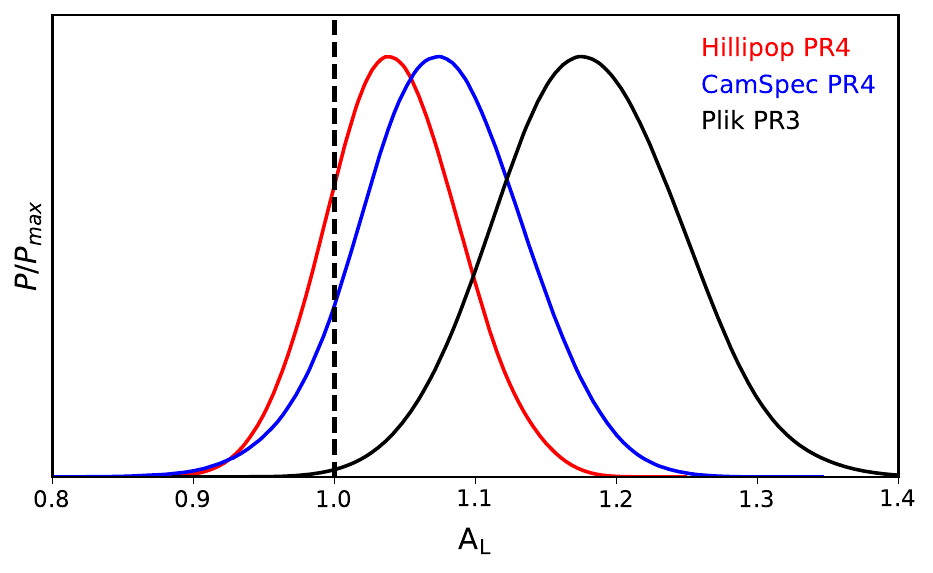}
    \caption{Posterior distributions for \Alens\ from \hillipop PR4, compared to \camspec (PR4) and \plik (PR3).}
    \label{fig:PR4_alens}
\end{figure}

In \citet{rosenberg:2022}, the \camspec likelihood associated with \lowl\ likelihoods from \planck 2018 also showed a decrease in the $\Alens$ parameter in \planck PR4 data compared to PR3 data, reducing the difference from unity from 2.4$\,\sigma$ to 1.7$\,\sigma$. When \lollipop is adopted as the \lowl\ polarized likelihood, instead of the \lowl\ likelihoods from \planck 2018, the constraint on \Alens\ from \camspec changed from $\Alens = 1.095 \pm 0.056$ to $\Alens = 1.075 \pm 0.058$, still a 1.3$\,\sigma$ difference from unity. We compare the posteriors for \plik (PR3), \camspec (PR4), and \hillipop (PR4) in Fig.~\ref{fig:PR4_alens}.
 
Previously, when there was a preference for $\Alens>1$, adding $\Alens$ as a seventh parameter could lead to shifts in other cosmological parameters \citep[e.g.,][]{planck2016-LI}. However, we confirm that with \hillipop on PR4, the \lcdm parameters are only affected through a very slight increase of the error bars, without significantly affecting the mean posterior values.

With the PR4 lensing reconstruction described in \citet{carron:2022}, the amplitude of the lensing power spectrum is $1.004 \pm 0.024$ relative to the \planck 2018 best-fit model. When combining CMB lensing with TTTEEE we then recover a tighter constraint on \Alens, with 
\begin{equation}
    \Alens = 1.037 \pm 0.037 \quad \text{(TTTEEE+lensing)}.
\end{equation}

\subsection{Curvature, \texorpdfstring{$\Omega_K$}{OmegaK}}
For the spatial curvature parameter, we report a significant difference with respect to \citet{planck2016-l06}, which used PR3 and reported a mild preference for closed models (i.e., $\Omega_K < 0$). Indeed, with \hillipop V4.2, the measurements are consistent with a flat universe ($\Omega_K = 0$) for all spectra.

\begin{figure}[ht!]
	\centering
	\includegraphics[width=0.9\columnwidth]{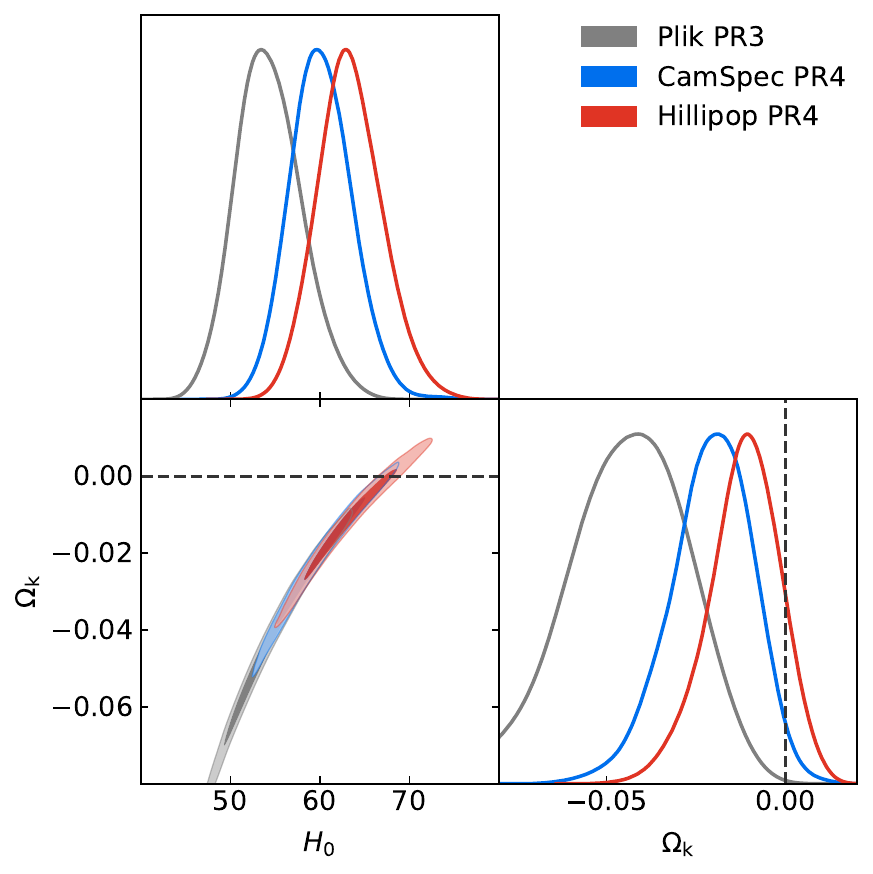}
	\caption{Posterior distributions in the $\Omega_{\rm K}$--$H_0$ plane using \hillipop PR4, compared to \camspec (PR4) and \plik (PR3).}
	\label{fig:PR4_PR3_Ok}
\end{figure}

As noticed in \citet{rosenberg:2022}, with \planck PR4, the constraint on $\Omegak$ is more precise and shifts toward zero, along the so-called geometrical degeneracy with $H_0$ (Fig.~\ref{fig:PR4_PR3_Ok}). 
Indeed, with \hillipop V4.2 on PR4, the posterior is more symmetrical and the mean value of the posterior for TTTEEE is 
\begin{equation}
\Omega_K = -0.012 \pm 0.010,
\end{equation}
which is only $1.2\,\sigma$ discrepant from zero.

This is to be compared to $\Omegak = -0.044_{-0.015}^{+0.018}$ obtained for \plik on PR3~\citep{planck2016-l06} and $\Omegak = -0.025_{-0.010}^{+0.013}$ obtained with \camspec on PR4~\citep{rosenberg:2022}. 

As a consequence, the tail of the 2-d posterior in the $H_0$--$\Omega_K$ plane at low $H_0$ and negative $\Omega_K$ is no longer favoured.
Indeed, when fitting for a non-flat Universe, the recovered value for the Hubble constant is $H_0 = (63.03 \pm 3.60)\,{\rm km}\,{\rm s}^{-1}\,{\rm Mpc}^{-1}$, only $1.3\,\sigma$ away from the constraint with fixed $\Omegak=0$.

\begin{figure}[ht!]
	\centering
	\includegraphics[width=0.9\columnwidth]{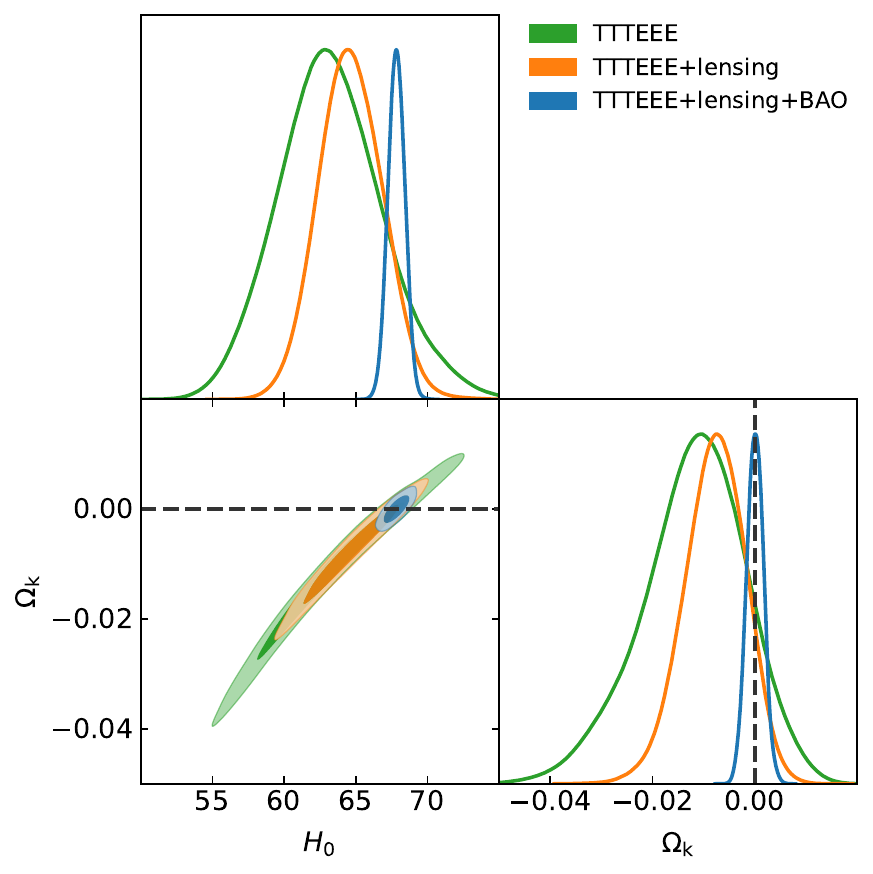}
	\caption{Posterior distributions in the $\Omega_K$--$H_0$ plane using \Planck\ PR4 TTTEEE (i.e., \lowlT+\lowlE+\hlp TTTEEE) in combination with lensing and BAO.}
	\label{fig:PR4_BAOlens_Ok}
\end{figure}

The combination of TTTEEE with lensing yields the improved constraint
\begin{equation}
    \Omega_K = -0.0078 \pm 0.0058 \quad \text{(TTTEEE+lensing)}.
\end{equation}
This is now compatible with the baryon acoustic oscillation measurements from SDSS, which are consistent with a flat Universe and give $\Omegak = -0.0022 \pm 0.0022$ \citep{alam:2021}.
Finally, the mean posterior for the combination of \planck\ PR4 TTTEEE with lensing and BAO is
\begin{equation}
    \Omega_K = 0.0000 \pm 0.0016 \quad \text{(TTTEEE+lensing+BAO)}.
\end{equation}
This is consistent with our Universe being spatially flat to within a 1$\,\sigma$ accuracy of 0.16\,\% (see Fig.~\ref{fig:PR4_BAOlens_Ok}).

\subsection{Effective number of relativistic species, \texorpdfstring{$N_\mathrm{eff}$}{Neff}}
\begin{figure}[!htbp]
	\centering
	\includegraphics[width=.85\columnwidth]{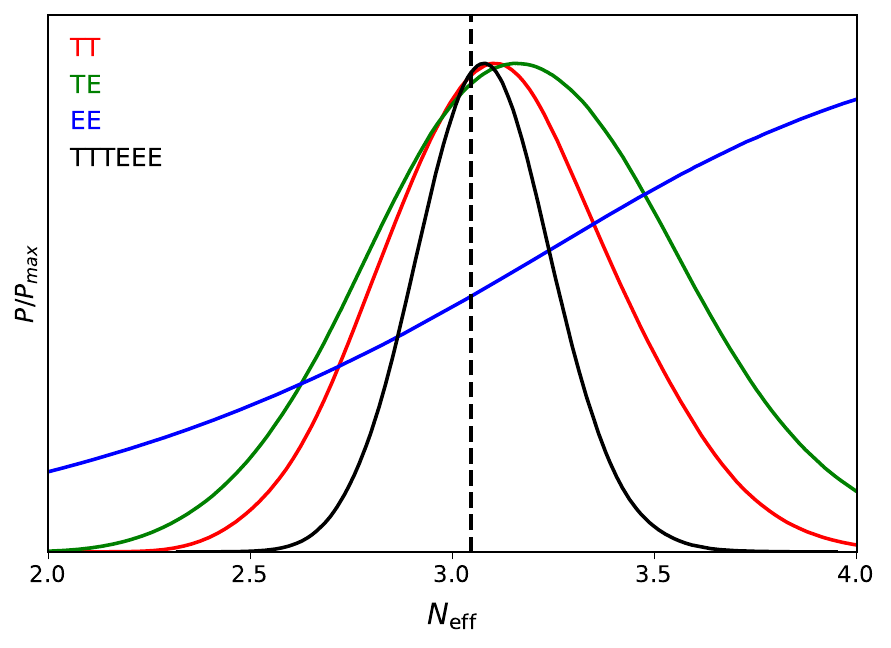}
	\caption{Posterior distributions for \Neff. The vertical dashed line shows the theoretical expectation ($\Neff = 3.044$).}
	\label{fig:PR4_Neff}
\end{figure}

Figure~\ref{fig:PR4_Neff} shows the posteriors for TT, TE, EE, and their combination when we consider the \Neff\ extension. Both TT and TE are compatible with similar uncertainties, while EE is not sensitive to \Neff.

The mean posterior for TTTEEE is
\begin{equation}
    \Neff = 3.08 \pm 0.17.
\end{equation}
The uncertainties are comparable to \planck 2018 results \citep[$\Neff = 2.92 \pm 0.19$,][]{planck2016-l06} with a slight shift toward higher values, closer to the theoretical expectation $\Neff = 3.044$ \citep{akita:2020,froustey:2020,bennett:2021}, which was also reported with \camspec analysis based on PR4 data \citep[$\Neff = 3.00 \pm 0.21$,][]{rosenberg:2022}.

\subsection{Sum of the neutrino masses, \texorpdfstring{$\sum m_\nu$}{Snu}}

\begin{figure}[!htbp]
	\centering
	\includegraphics[width=.85\columnwidth]{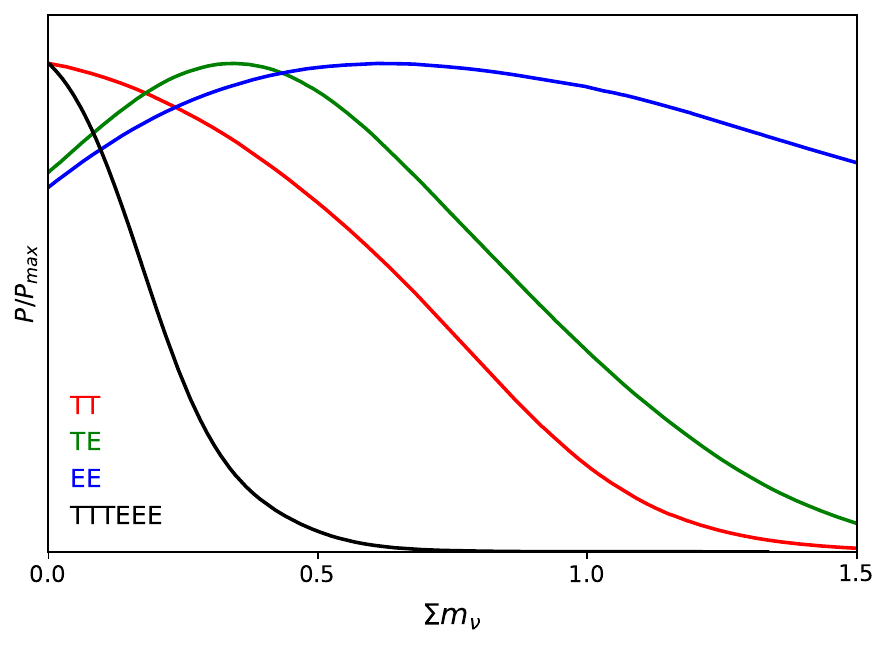}
	\caption{Posterior distributions for \mnu. Units are electronvolts.}
	\label{fig:PR4_Mnu}
\end{figure}

Figure~\ref{fig:PR4_Mnu} shows the posterior distribution for the sum of the neutrino masses, \mnu. There is no detection of the effects of neutrino mass and we report an upper limit of
\begin{equation}
    \mnu < 0.39\,\text{eV} \quad \text{(95\,\% CL, TTTEEE)}.
\end{equation}
Despite the increase in sensitivity associated with PR4, the constraint is slightly weaker (the upper limit is larger) than the one reported for \planck 2018: $\mnu < 0.26$\,eV at 95\,\% CL. 
Our constraint is comparable to \camspec, which gives $\mnu < 0.36$\,eV at 95\,\% CL.

As explained in \citet{couchot_mnu:2017} and \citet{planck2016-l06}, this is directly related to the value of \Alens. Indeed, the correlation between \Alens\ and \mnu\ pushes the peak posterior of \mnu\ toward negative values when \Alens\ is fixed to unity; the data, however, prefer values of \Alens\ larger than 1. With \hillipop V4.2, the value of \Alens\ reported in this work is more compatible with unity ($\Alens = 1.039 \pm 0.052$, see Sect.~\ref{sec:alens}), thus, the posterior for \mnu\ is shifted to higher values, with a peak closer to zero, increasing the upper limit accordingly.

\begin{figure}[ht!]
	\centering
	\includegraphics[width=0.9\columnwidth]{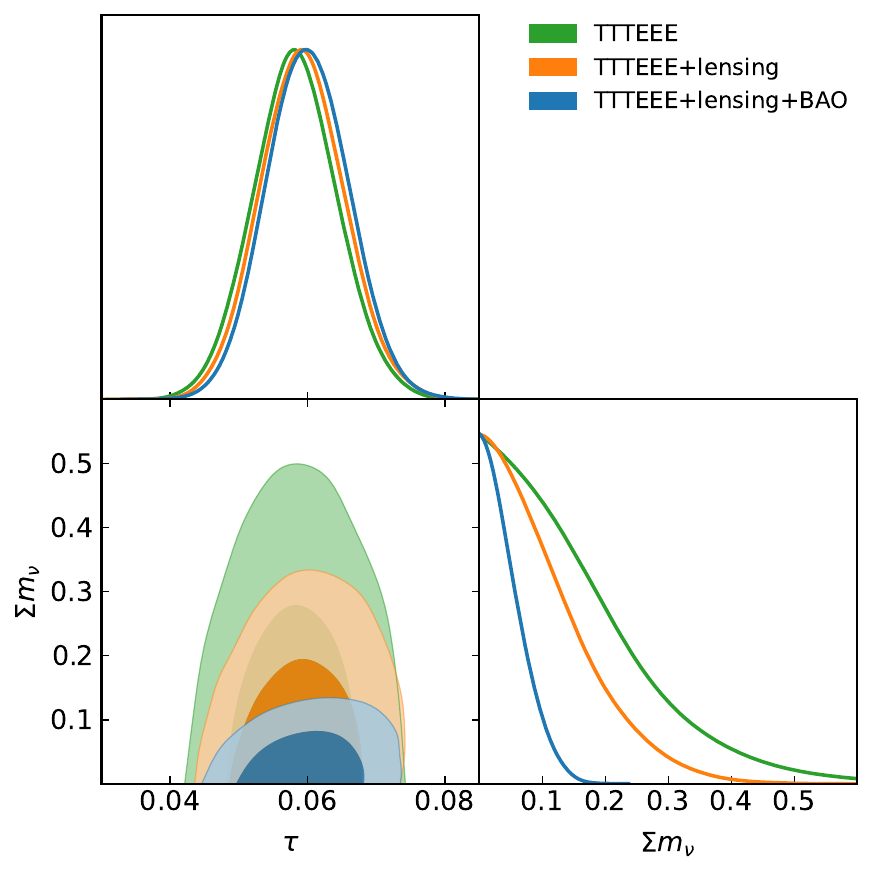}
	\caption{Posterior distributions in the $\mnu$--$\tau$ plane using \Planck\ PR4 TTTEEE (i.e., \lowlT+\lowlE+\hlp TTTEEE) in combination with lensing and BAO.}
	\label{fig:PR4_BAOlens_Mnu}
\end{figure}

Figure~\ref{fig:PR4_BAOlens_Mnu} shows constraints in the \mnu--$\tau$ plane when combining our new likelihood with with CMB lensing and BAO data. This combination further strengthens the limits to
\begin{eqnarray}
    {\textstyle \sum} m_\nu &<& 0.26\,\text{eV} \quad \text{(95\,\% CL, TTTEEE+lensing)},\\
    {\textstyle \sum} m_\nu &<& 0.11\,\text{eV} \quad \text{(95\,\% CL, TTTEEE+lensing+BAO)}.   
\end{eqnarray}
This is slightly tighter than the upper limit from \planck 2018 ($\mnu < 0.12$\,eV) and getting close to the lower-limit for the inverted mass hierarchy \citep[$\mnu \ga 0.1$\,eV, see e.g.,][]{Jimenez:2022}.

\section{Conclusions}
In this paper, we have derived cosmological constraints using CMB anisotropies from the final \planck data release (PR4).
We detailed a new version of a CMB \mbox{\highl} likelihood based on cross-power spectra computed from the PR4 maps. This version of \hillipop, labelled V4.2, uses more sky (75\,\%) and a wider range of multipoles.
Our likelihood makes use of physically-motivated models for foreground-emission residuals. Using only priors on the foreground spectral energy distributions, we found amplitudes for residuals consistent with expectations.
Moreover, we have shown that the impact of this modelling on cosmological \lcdm parameters is negligible.

Combined with the \lowl\ $EE$ likelihood \lollipop, we derived constraints on \lcdm and find good consistency with \planck 2018 results (based on PR3) with better goodness-of-fit and higher sensitivity (from 10\,\% to 20\,\%, depending on the parameters). In particular, we now constrain the reionization optical depth at the 10\,\% level. We found a value for the Hubble constant consistent with previous CMB measurements and thus still in tension with distance-ladder results. 
We also obtained a lower value for $S_8$, alleviating the CMB versus large-scale structure tension to 1.5$\,\sigma$.

We found good consistency with the other published CMB likelihood analysis based on PR4, \camspec \citep{rosenberg:2022}, which relies on a procedure to clean power spectra prior to constructing the likelihood. The consistency of the results using two different approaches reinforces the robustness of the results obtained with \planck data.

We also add constraints from PR4 lensing, making the combination the most constraining data set that is currently available from \planck. Additionally we explore adding baryon acoustic oscillation data, which tightens limits on some particular extensions to the standard cosmology.

We provided constraints on some extensions to \lcdm, including the lensing amplitude \Alens, the curvature $\Omega_K$, the effective number of relativistic species \Neff, and the sum of the neutrino masses \mnu. 
For both \Alens\ and \Omegak, our results show a significant reduction of the so-called ``tensions'' with standard \lcdm, together with a reduction of the uncertainties. The final constraints indeed are fully compatible with \lcdm predictions. 
In particular, with the new version of the likelihood presented in this work, we report $\Alens = 1.039 \pm 0.052$, entirely compatible with the \lcdm prediction.
The better agreement is explained both by the improvement of the \planck maps thanks to the \npipe processing (with less noise and better systematic control in polarization) and the use of the \lollipop and \hillipop likelihoods.
\\

\begin{acknowledgements}
\planck is a project of the European Space Agency (ESA) with instruments provided by two scientific consortia funded by ESA member states and led by Principal Investigators from France and Italy, telescope reflectors provided through a collaboration between ESA and a scientific consortium led and funded by Denmark, and additional contributions from NASA (USA).
Some of the results in this paper have been derived using the {\tt HEALPix} package. 
We acknowledge use of the following packages: \texttt{xQML}, for the computation of large-scale power spectra (\myurl{gitlab.in2p3.fr/xQML}); \xpol, for the computation of large-scale power spectra (\myurl{gitlab.in2p3.fr/tristram/Xpol});
\texttt{Cobaya}, for the sampling of the likelihoods (\myurl{github.com/CobayaSampler}); and CLASS (\myurl{github.com/lesgourg/class\_public}) and CAMB (\myurl{github.com/cmbant/CAMB}) for calculating power spectra.
We gratefully acknowledge support from the CNRS/IN2P3 Computing Center for providing computing and data-processing resources needed for this work.
This research was enabled in part by support provided by the Digital Research Alliance of Canada (\myurl{alliancecan.ca)}.
This project has received funding from the European Research Council (ERC) under the European Union’s Horizon 2020 research and innovation programme (grant agreement No 788212).
\\

\noindent \textit{Data availability.} The \planck PR4 data are publicly available on the \planck Legacy Archive (\myurl{pla.esac.esa.int}). Both likelihoods \lollipop and \hillipop based on PR4 are publicly available on github (\myurl{github.com/planck-npipe}) as external likelihoods for \texttt{Cobaya}.
\end{acknowledgements}

\bibliographystyle{aat}
\bibliography{Planck_bib,hillipop}

\onecolumn
\appendix

\section{Foregrounds and instrumental parameters}
\label{app:priors}
Here we describe the ``nuisance'' parameters relating to foreground emission components and the instrument. They are listed in Table~\ref{tab:params} together with their prior and the recovered best-fit value for the combination TTTEEE.

\vspace{-0.5cm}
\begin{table}[!htbp]
\begin{center}
\begin{tabular}{llcc}
\hline
\hline
\noalign{\vskip 1pt}
Name & Definition & Prior & Mean \\
\hline
\noalign{\vskip 1pt}
$A_\mathrm{planck}$ & Absolute calibration  & $1.0000 \pm 0.0025$ & $0.9997 \pm 0.0024$\\
$c_\mathrm{100A}$ & Map recalibration (100A)  & [0.9,1.1] & $1.003 \pm 0.007$\\
$c_\mathrm{100B}$ & Map recalibration (100B)  & [0.9,1.1] & $1.004 \pm 0.007$\\
$c_\mathrm{143A}$ & Map recalibration (143A)  & 1.0 (fixed) & \\
$c_\mathrm{143B}$ & Map recalibration (143B)  & [0.9,1.1] & $1.004 \pm 0.006$\\
$c_\mathrm{217A}$ & Map recalibration (217A)  & [0.9,1.1] & $1.001 \pm 0.008$\\
$c_\mathrm{217B}$ & Map recalibration (217B)  & [0.9,1.1] & $1.001 \pm 0.008$\\
$\eta_{\rm 100-A}$ & Cross-polarization (100-A) & $1.000$ (fixed) \\  
$\eta_{\rm 100-B}$ & Cross-polarization (100-B) & $1.000$ (fixed) \\  
$\eta_{\rm 143-A}$ & Cross-polarization (143-A) & $1.000$ (fixed) \\  
$\eta_{\rm 143-B}$ & Cross-polarization (143-B) & $1.000$ (fixed) \\  
$\eta_{\rm 217-A}$ & Cross-polarization (217-A) & $0.975$ (fixed) \\  
$\eta_{\rm 217-B}$ & Cross-polarization (217-B) & $0.975$ (fixed) \\  
\hline
\noalign{\vskip 2pt}
$c_\mathrm{dust}^{T}$ & Rescaling for Galactic dust in temperature  & $1.0 \pm 0.1$ & $ 1.08 \pm  0.03$\\
    \noalign{\vskip 1pt}
$c_\mathrm{dust}^{P}$ & Rescaling for Galactic dust in polarization  & $1.0 \pm 0.1$ & $ 1.20 \pm  0.03$\\
$A_\mathrm{radio}$ & Amplitude for radio sources  & [0,150] & $ 63.3 \pm 4.7$\\
$A_\mathrm{IR}$ & Amplitude for IR sources  & [0,150] & $  6.07 \pm   0.63$\\
$A_\mathrm{CIB}$ & Amplitude for the CIB  & [0,20] & $ 1.03 \pm  0.34$\\
$A_\mathrm{tSZ}$ & Amplitude for the tSZ effect  & [0,50] & $ 5.87 \pm  1.66$\\
$A_\mathrm{kSZ}$ & Amplitude for the kSZ effect  & [0,50] & $<7.6$\\
$\xi_\mathrm{SZ\times CIB}$ & Cross-correlation tSZ$\times$CIB  & [$-$1,1] & $ 0.46 \pm  0.30$\\
\noalign{\vskip 1pt}
\hline
\noalign{\vskip 2pt}
$\beta_\mathrm{dust}^T$ & Spectral index for dust in temperature  & $1.51 \pm 0.01$ & $ 1.51 \pm  0.01$\\
\noalign{\vskip 1pt}
$\beta_\mathrm{dust}^P$ & Spectral index for dust in polarization  & $1.59 \pm 0.02$ & $ 1.59 \pm  0.02$\\
$\beta_\mathrm{CIB}$ & Spectral index for CIB  & $1.75 \pm 0.06$ & $ 1.85 \pm  0.06$\\
$\beta_\mathrm{radio}$ & Spectral index for radio sources  & $-$0.8 & \\
\hline
\end{tabular}
\caption{Instrumental and foreground parameters for the \hillipop likelihood with their respective priors. 
Amplitudes refer to $D_\ell = \ell(\ell+1)C_\ell /2\pi$ for $\ell=3000$ at 143\GHz, except for dust coefficients, $c_\mathrm{dust}$, for which the priors are found by rescaling the dust power spectrum at 353\GHz.}
\label{tab:params}
\end{center}
\end{table}

\vspace{-1cm}

\section{Best-fit model components}
\label{app:components}
\begin{figure*}[!htbp]
	\centering
	\includegraphics[height=140px]{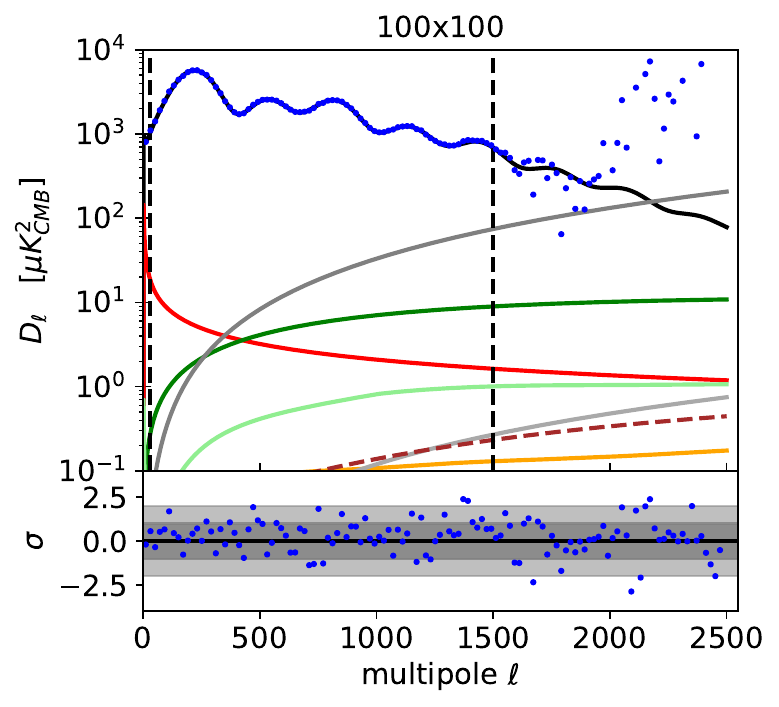}
	\includegraphics[height=140px]{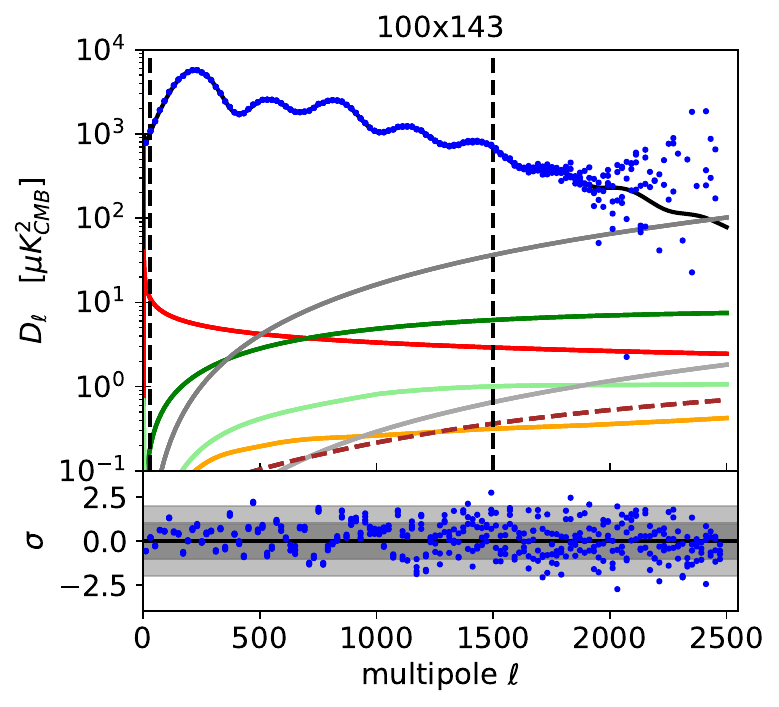}
	\includegraphics[height=140px]{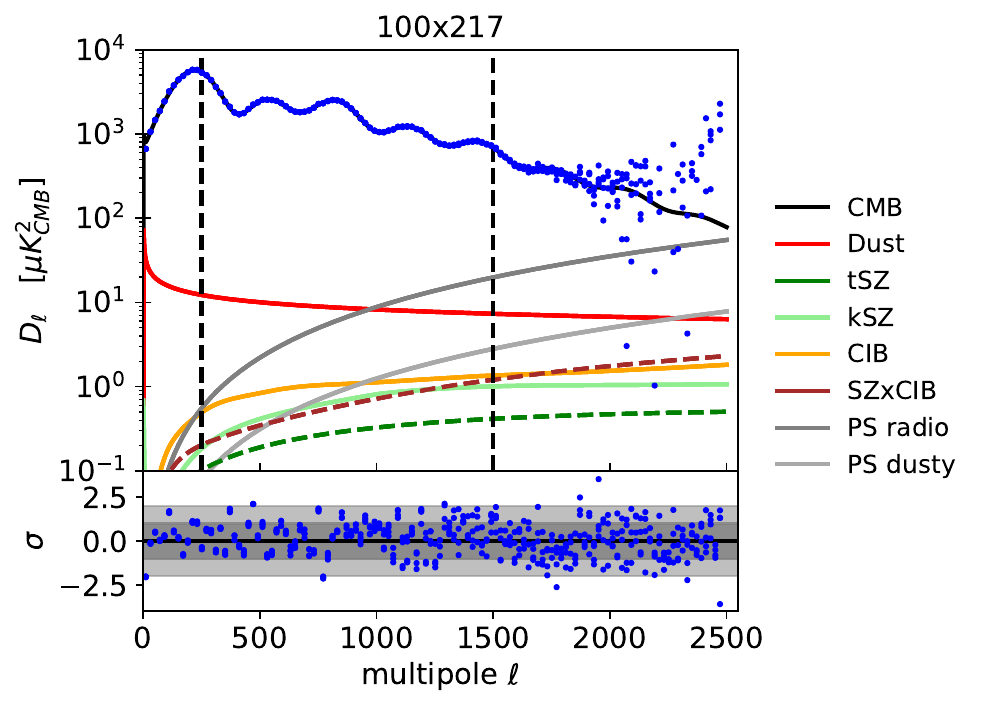}\\
	\includegraphics[height=140px]{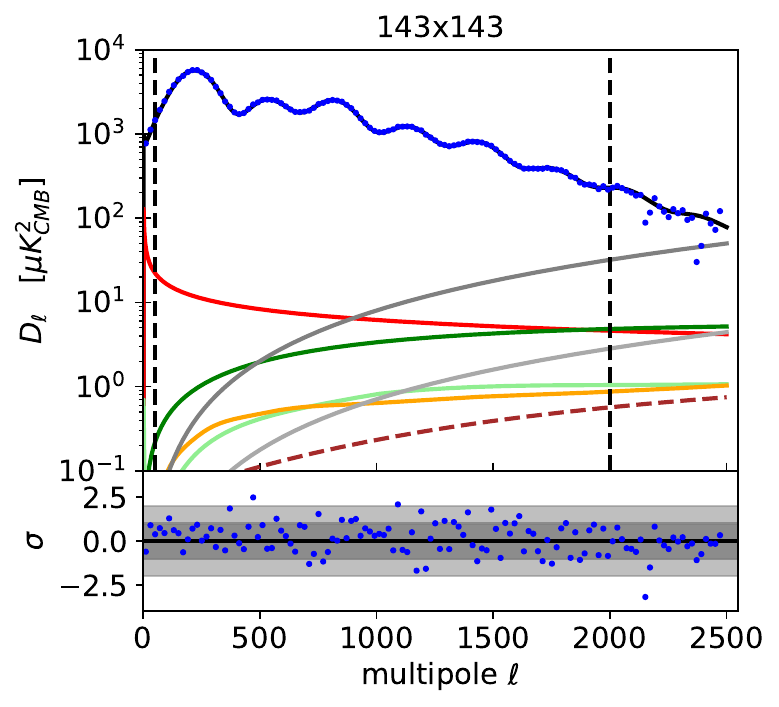}
	\includegraphics[height=140px]{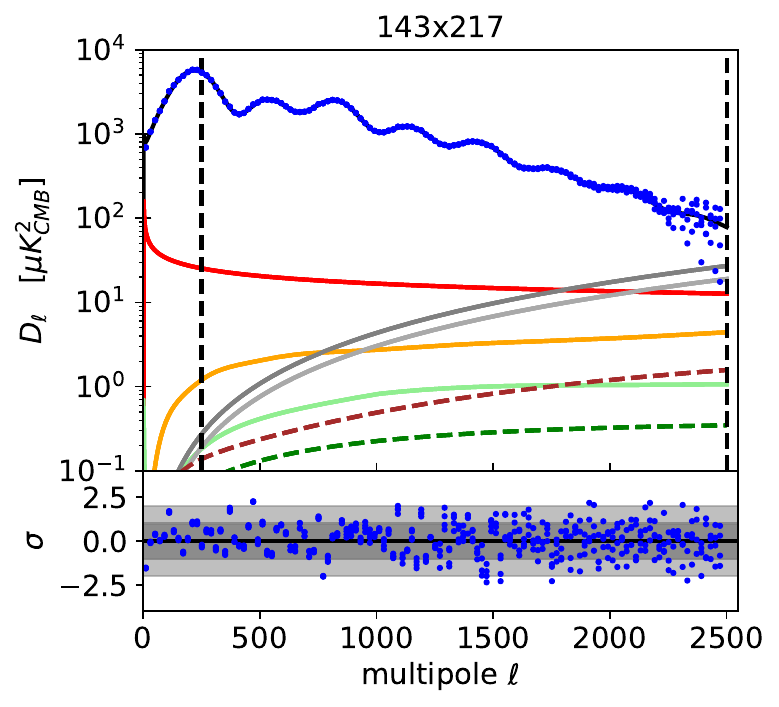}
	\includegraphics[height=140px]{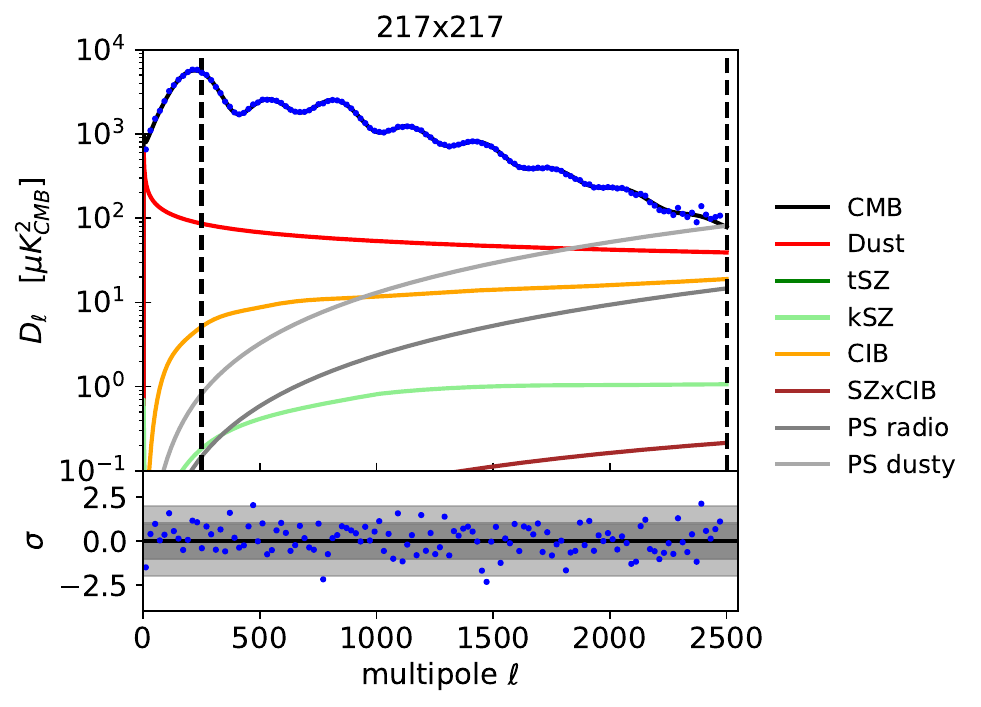}
	\caption{Best-fit model for each cross-frequency power spectrum in temperature, including emission from CMB, dust, tSZ, kSZ, CIB, SZ$\times$CIB, and Poisson-noise from radio sources and dusty galaxies. Negative components are shown as dashed lines. Vertical black dashed lines show the range of multipoles considered in \hillipop V4.2. The bottom panels show the residuals normalized by the error bars. Data are binned with $\Delta\ell=20$ for this plot.}
	\label{fig:TT_components}
\end{figure*}

Here we present our results for the best-fitting model components for each cross-power spectrum.
These are shown in Fig.~\ref{fig:TT_components} and the corresponding $\chi^2$ values are given in Table~\ref{tab:chi2_spectra}.

\begin{table}[ht!]
    \begin{center}
    \begin{tabular}{lcccccccc}
    \hline
    \hline
    \noalign{\vskip 1pt}
    Cross-spectrum     & \multicolumn{2}{c}{$TT$} &  \multicolumn{2}{c}{$EE$} &  \multicolumn{2}{c}{$TE$}  & \multicolumn{2}{c}{$ET$}     \\
         &  \hspace{.5em}$\chi^2/n_{\rm d}$   & $\delta \sigma(\chi^2)$\hspace{.5em}
         &  \hspace{.5em}$\chi^2/n_{\rm d}$   & $\delta \sigma(\chi^2)$\hspace{.5em}
         &  \hspace{.5em}$\chi^2/n_{\rm d}$   & $\delta \sigma(\chi^2)$\hspace{.5em}   
         &  \hspace{.5em}$\chi^2/n_{\rm d}$   & $\delta \sigma(\chi^2)$\hspace{.5em}   \\
    \noalign{\vskip 1pt}
    \hline
    \noalign{\vskip 1pt}
100A$\times$100B  & 1590.0 / 1471 & 2.19 & 1079.1 / 1101 & $-$0.47 & 1597.4 / 1471 & \phantom{$-$}2.33 & 1450.1 / 1471 & $-$0.39 \\
\hline
\noalign{\vskip 1pt}
100A$\times$143A  & 1616.5 / 1471 & 2.68 & 1551.5 / 1471 & \phantom{$-$}1.48 & 1564.5 / 1471 & \phantom{$-$}1.72 & 1490.8 / 1471 & \phantom{$-$}0.37 \\
100A$\times$143B  & 1605.1 / 1471 & 2.47 & 1431.3 / 1471 & $-$0.73 & 1396.4 / 1471 & $-$1.38 & 1520.6 / 1471 & \phantom{$-$}0.92 \\
100B$\times$143A  & 1596.3 / 1471 & 2.31 & 1485.7 / 1471 & \phantom{$-$}0.27 & 1535.2 / 1471 & \phantom{$-$}1.18 & 1615.8 / 1471 & \phantom{$-$}2.67 \\
100B$\times$143B  & 1576.5 / 1471 & 1.94 & 1495.5 / 1471 & \phantom{$-$}0.45 & 1466.9 / 1471 & $-$0.08 & 1614.6 / 1471 & \phantom{$-$}2.65 \\
\hline
\noalign{\vskip 1pt}
100A$\times$217A  & 1379.1 / 1251 & 2.56 & 1331.5 / 1251 & \phantom{$-$}1.61 & 1478.0 / 1401 & \phantom{$-$}1.45 & 1432.3 / 1401 & \phantom{$-$}0.59 \\
100A$\times$217B  & 1364.5 / 1251 & 2.27 & 1278.4 / 1251 & \phantom{$-$}0.55 & 1481.3 / 1401 & \phantom{$-$}1.52 & 1445.3 / 1401 & \phantom{$-$}0.84 \\
100B$\times$217A  & 1336.8 / 1251 & 1.71 & 1283.0 / 1251 & \phantom{$-$}0.64 & 1507.3 / 1401 & \phantom{$-$}2.01 & 1545.9 / 1401 & \phantom{$-$}2.74 \\
100B$\times$217B  & 1335.0 / 1251 & 1.68 & 1218.3 / 1251 & $-$0.65 & 1466.8 / 1401 & \phantom{$-$}1.24 & 1505.7 / 1401 & \phantom{$-$}1.98 \\
\hline
\noalign{\vskip 1pt}
143A$\times$143B  & 2108.5 / 1951 & 2.52 & 1995.3 / 1971 & \phantom{$-$}0.39 & 2014.7 / 1971 & \phantom{$-$}0.70 & 1972.1 / 1971 & \phantom{$-$}0.02 \\
\hline
\noalign{\vskip 1pt}
143A$\times$217A  & 2324.1 / 2251 & 1.09 & 1647.4 / 1751 & $-$1.75 & 1847.9 / 1801 & \phantom{$-$}0.78 & 1868.7 / 1801 & \phantom{$-$}1.13 \\
143A$\times$217B  & 2327.3 / 2251 & 1.14 & 1853.6 / 1751 & \phantom{$-$}1.73 & 1746.9 / 1801 & $-$0.90 & 1898.1 / 1801 & \phantom{$-$}1.62 \\
143B$\times$217A  & 2351.2 / 2251 & 1.49 & 1725.3 / 1751 & $-$0.43 & 1812.0 / 1801 & \phantom{$-$}0.18 & 1835.4 / 1801 & \phantom{$-$}0.57 \\
143B$\times$217B  & 2321.0 / 2251 & 1.04 & 1799.5 / 1751 & \phantom{$-$}0.82 & 1696.4 / 1801 & $-$1.74 & 1862.6 / 1801 & \phantom{$-$}1.03 \\
\hline
\noalign{\vskip 1pt}
217A$\times$217B  & 2283.6 / 2251 & 0.49 & 1732.8 / 1751 & $-$0.31 & 1625.4 / 1701 & $-$1.30 & 1725.8 / 1701 & \phantom{$-$}0.43 \\
\hline
    \end{tabular}
    \caption{$\chi^2$ values for each cross-spectrum compared to the size of the data vector ($n_{\rm d}$).}
    \label{tab:chi2_spectra}
    \end{center}
\end{table}

\end{document}